\documentclass[12pt]{article}
\pdfoutput=1

\usepackage[nosort]{cite}
\usepackage{epsfig}
\usepackage{amsmath}
\usepackage{amsfonts}
\usepackage{amssymb}
\usepackage{multirow}
\usepackage{array}

\newcommand{\dslash}{\not\!\partial}
\newcommand{\Dslash}{\not\!\! D}
\newcommand{\eps}{\varepsilon}
\newcommand{\vphi}{\varphi}

 \def    \ptj           {\mbox{$p_{T j}$}}
 \def    \etaj           {\mbox{$\eta_{ j}$}}
 \def    \ptl           {\mbox{$p_{T l}$}}
 \def    \etal           {\mbox{$\eta_{ l}$}}
 \def    \gev            {\mbox{$\mathrm{GeV}$}}
\newcommand{\tev}{\mbox{$\mathrm{TeV}$}}
\newcommand{\etmiss}{ \not \hskip -4pt E_T}
\newcommand{\nottop}{{\!\not\, t}}

\newcommand{\Scal}{\mathcal{S}}
\newcommand{\Bcal}{\mathcal{B}}

\begin{document}

\setcounter{page}{0}
\thispagestyle{empty}

%%%%%%%%%%%%%%%%%%%%%%%%%%%%%%%%%%%%%%%%%%%%%%%%%%%%%%%%%%%%%%%%%%%%%%%%%%%%%%%

\vspace*{0.8cm}

\begin{center}
{\bf \Large {
Heavy-light decay topologies as a new 
strategy  \\[10pt] to discover a heavy gluon
}}
\end{center}

\vskip 20pt

\begin{center}
{\large Cesare Bini$^a$, Roberto Contino$^a$, Natascia Vignaroli$^{a,b}$} 
\end{center}

\vskip 20pt

\begin{center}
$^{a}\,${\it Dipartimento di Fisica, Sapienza Universit\`a di Roma and INFN, Italy} \\[0.3cm]
$^b\,${\it Department of Physics and Astronomy, Iowa State University, Ames, IA 50011}
\end{center}

\vskip 50pt

\begin{abstract}
\vskip 3pt
\noindent
We study the collider phenomenology of the lightest Kaluza-Klein excitation of the gluon, $G^*$, in 
theories with a warped extra dimension. We do so  by means of a two-site effective lagrangian which 
includes only the lowest-lying spin-1 and spin-1/2 resonances.
We point out the importance of the decays of $G^*$ to one SM plus one heavy fermion, that were overlooked in 
the previous literature. It turns out that, when kinematically allowed,  such heavy-light decays are powerful channels 
for discovering the $G^*$. In particular, we present a parton-level Montecarlo analysis of the final state $Wtb$ that 
follows from the decay of $G^*$ to one SM top or bottom quark plus its heavy partner. We find that at $\sqrt{s} = 7\,$TeV 
and with $10\,\text{fb}^{-1}$ of integrated luminosity, the LHC can discover  a KK gluon with mass in the range 
$M_{G^*} = (1.8 - 2.2)\,$TeV if its coupling to a pair of light quarks is $g_{G^*q\bar q} = (0.2-0.5) g_3$.
The same process is also competitive for the discovery of the top and bottom partners as well.
We find, for example, that the LHC  at $\sqrt{s} = 7\,$TeV  can discover a 1 TeV KK bottom quark with an integrated
luminosity of $(5.3 - 0.61) \,\text{fb}^{-1}$ for $g_{G^*q\bar q} = (0.2-0.5) g_3$.
\end{abstract}

\vskip 13pt
\newpage

%%%%%%%%%%%%%%%%%%%%%%%%%%%%%%%%%%%%%%%%%%%%%%%%%%%%%%%%%%%%%%%%%%%%%%%%%%%%%

%%%%%%%%%%%%%%%%%%%%%%%%%%%%%%%%%%%%%%
\section{Introduction}
\label{sec:introduction}
%%%%%%%%%%%%%%%%%%%%%%%%%%%%%%%%%%%%%%

One of the robust predictions of theories with a warped extra-dimension~\cite{Randall:1999ee}
and fields propagating in the bulk~\cite{bulkgauge,Grossman:1999ra,Gherghetta:2000qt} is the existence 
of heavier copies of the gluon with $\sim\,$TeV mass, its  Kaluza-Klein (KK) excitations.
The phenomenology of the lightest of the KK gluons, which we will denote as $G^*$, has been extensively 
studied~\cite{Agashe:2006hk,Lillie:2007yh,Lillie:2007ve,Djouadi:2007eg,Frederix:2007gi,Redi:2011zi}.
In the theoretically attractive framework where the hierarchy of masses and mixings  of the Standard Model (SM) quarks follows from 
the geography of  wave functions in the bulk, while the 5D Yukawa couplings are flavor 
anarchic~\cite{Grossman:1999ra,Gherghetta:2000qt,Huber:2000ie}, the couplings
of the heavy gluon $G^*$ to the SM quarks are dictated by an inverse hierarchy, where the largest coupling is to the top quark.
In this case the decay rate of $G^*$ to pairs of top quarks is typically large and dominates over the decay rate to two light quarks.

Experiments at the Tevatron and the LHC have been testing this theoretical picture performing several searches for the heavy gluon $G^*$
in dijet final states~\cite{Chatrchyan:2011ns,Aad:2011fq} and in the $t\bar t$ 
channel~\cite{CDF9164,D05882,ATLAS087,ATLAS123,CMS055,CMS006}.~\footnote{Here we list only those experimental searches that presently
impose the strongest bounds, omitting previous or less constraining analyses.}  
In the latter case, some analyses do not attempt to reconstruct the top quarks,
and are in fact searches for $WbWb$ resonances. Others instead look for highly boosted top quarks and make use of dedicated top-tagging
techniques to reduce the SM background, following the original suggestion of Refs.~\cite{Agashe:2006hk,Lillie:2007yh}.
Taken all together, these experimental results impose important constraints on the value of the mass and couplings of the heavy gluon,
and in turn on the parameter space of the extra-dimensional theories.

In addition to the KK gluon, new heavy fermions are also naturally predicted in theories with a warped extra dimension 
as the Kaluza-Klein excitations of the SM quarks and leptons. In particular, in 
models where  the Higgs doublet is a pseudo Nambu-Goldstone
boson of a larger symmetry breaking~\cite{Contino:2003ve,Agashe:2004rs}, the  lightest KK top regulates the quadratically divergent 
contribution of the SM top quark to the Higgs mass term.  By naturalness arguments,
it is thus expected to have a mass below $\sim 1\, $TeV.
The coupling of $G^*$ to the heavy fermions is typically large, so that, if kinematically allowed, the heavy gluon decays to a pair of 
heavy fermions with a large rate. In this limit the total decay width of  $G^*$ becomes
very large, due both to the large value of the couplings and to the multiplicity of available decay channels, making its discovery 
quite challenging~\cite{Carena:2007tn}.

On the other hand, $G^*$ can also decay to one SM plus one heavy fermion. Specifically, the most important channels in scenarios with
flavor anarchy are those with one SM top or bottom quark plus its heavy copy.
If the threshold to two heavy fermions is kinematically closed, the branching ratio of these heavy-light channels
can be large in a wide portion of the parameter space. At the same time the branching ratio to pairs of SM quarks is reduced,
making the standard analyses of dijet and $t\bar t$ final states less constraining.
The importance of the heavy-light decays at the Tevatron  was already pointed out in Ref.~\cite{Dobrescu:2009vz} in the context
of a different theoretical framework. 
No detailed study however has been done so far, to our knowledge, to assess the
prospects of observing the heavy-light decays of $G^*$ at the LHC.

In this paper we present a first analysis of this kind and outline a simple cut-based strategy to 
maximize the signal significance over the SM background.
Some preliminary results have already been presented in Ref.~\cite{Vignaroli:2011ik}.
We find that, thanks to their distinct topology, decays to one SM plus one heavy fermion are extremely clean channels
with a strong potential for the discovery  not just of $G^*$, but of the KK top and bottom as well.
Our analysis suggests that the discovery reach on these latter can be 
larger than the one obtained by exploiting the more studied processes of QCD pair 
production~\cite{AguilarSaavedra:2005pv,Carena:2007tn,Skiba:2007fw,Contino:2008hi,AguilarSaavedra:2009es} 
and electroweak single production~\cite{Azuelos:2004dm,Atre:2008iu,Mrazek:2009yu,Atre:2011ae}.

Although our study is performed in the specific context of warped extra-dimensional theories, 
we believe that our results apply to a broader class of models beyond the SM where both a heavy spin-1 color-octect
and heavy top or bottom fermions exist, like for example the top-quark seesaw model~\cite{Dobrescu:1997nm,Chivukula:1998wd}.

The paper is organized as follows. In Section~\ref{sec:model} we define the effective two-site model that we adopt
to study the phenomenology of the heavy gluon and heavy fermions. We derive the $G^*$ production cross section and
the relevant decay rates, showing that the branching ratio of $G^*$ to one SM plus one heavy fermion is large when kinematically
allowed. In Section~\ref{sec:analysis} we study the prospects of observing the heavy-light decays of $G^*$ in $Wtb$ events at the LHC.
We perform a Montecarlo simulation of the signal and the main SM backgrounds and design a simple set of kinematic cuts
that maximizes the discovery significance. In Section~\ref{sec:conclusions} we discuss the results obtained and draw our conclusions.
Finally, Appendix~\ref{app:Lagrangian} contains a few results used in discussing the effective model, while details on how the statistical 
errors are computed  are given  in Appendix~\ref{app:errors}.

%%%%%%%%%%%%%%%%%%%%%%%%%%%%%%%%%%%%%%
\section{An effective theory of the heavy gluon and the top partners}
\label{sec:model}
%%%%%%%%%%%%%%%%%%%%%%%%%%%%%%%%%%%%%%

Instead of considering the full set of particles and interactions arising in a specific model, we will work in the framework of
an effective theory that includes only the lowest-lying resonances and reproduces, in the same spirit of  chiral lagrangians in QCD, 
the low-energy limit of a large set of warped extra-dimensional theories with a custodial symmetry in the bulk~\cite{Agashe:2003zs}. 
Specifically, we will construct the effective lagrangian by following the rules of deconstruction defined in Ref.~\cite{Contino:2006nn},
and adopt the `dual' language of strongly-interacting dynamics to describe the theory~\cite{AdSCFT,holography}.
In this perspective, the new heavy particles arise as composite states of a new strong sector, which are coupled
to the SM elementary ones via linear mixing terms, leading to a scenario of partial compositeness~\cite{Kaplan:1991dc,Contino:2006nn}. 
The Higgs is also a bound state of the new dynamics and has direct couplings only to the composite fermions.

\subsection{The model}

We consider the case in which the composite sector has
an $SU(3)_c \times O(4) \times U(1)_X$ global symmetry, with $O(4) \supset SO(4) \sim SU(2)_L \times SU(2)_R$, where $O(4)$ contains
a discrete parity $P_{LR}$ under which $SU(2)_L$ and $SU(2)_R$ are exchanged.
The two building blocks of the model are the elementary sector and the composite sector. The particle content of the elementary sector
is that of the SM without the Higgs, and the $SU(3)_c \times SU(2)_L \times U(1)_Y$ elementary fields gauge the corresponding global
invariance of the strong dynamics, with $Y = T_{R}^3 + X$.
The composite sector comprises a heavy gluon, $G^*$, which transforms as $(\mathbf{8}, \mathbf{1},\mathbf{1})_0$ under 
$SU(3)_c \times  SU(2)_L \times SU(2)_R \times U(1)_X$, the composite Higgs
\begin{equation} \label{eq:higgs}
{\cal H} = (\mathbf{1},\mathbf{2},\mathbf{2})_{0} = 
 \begin{bmatrix} \phi_0^\dagger & \phi^+ \\ - \phi^- & \phi_0 \end{bmatrix} \, ,
\end{equation}
and the following set of vector-like composite fermions:
\begin{equation} \label{eq:fermions}
\begin{aligned}
{\cal Q} & = (\mathbf{3},\mathbf{2},\mathbf{2})_{2/3} = \begin{bmatrix} T & T_{5/3} \\ B & T_{2/3} \end{bmatrix} \, , \qquad
& \tilde T & = (\mathbf{3},\mathbf{1},\mathbf{1})_{2/3}\, ,  \\[0.5cm]
{\cal Q^\prime} & = (\mathbf{3},\mathbf{2},\mathbf{2})^\prime_{-1/3} = \begin{bmatrix} B_{-1/3} & T^\prime \\ B_{-4/3} & B^\prime \end{bmatrix} \, , \qquad
& \tilde B &= (\mathbf{3},\mathbf{1},\mathbf{1})_{-1/3}\, .
\end{aligned}
\end{equation}
Other composite states, for example  spin-1 resonances with electroweak quantum numbers, are in general present in the spectrum of realistic models, 
but will not be considered here for simplicity. The quantum numbers of the composite fermions and the Higgs under 
$SU(3)_c \times  SU(2)_L \times SU(2)_R \times U(1)_X$ are those specified in  eqs.(\ref{eq:higgs}),(\ref{eq:fermions}).
The fermions, in particular, can be arranged in two fundamentals of $SO(5)$, and in fact our effective model describes the low-energy 
regime of the $SO(5)/SO(4)$ theories introduced in Refs.~\cite{Contino:2006qr,Carena:2006bn}, in the limit in which only the leading terms in 
an expansion in powers of the Higgs field are retained (see Refs.~\cite{Panico:2011pw,DeCurtis:2011yx}, for two- and three-site effective theories 
where the full Higgs non-linearities are included). For this reason, we will dub our two-site model as TS5. As it will be even more evident shortly, it 
extends the two-site model of  Ref.~\cite{Contino:2008hi} to include the heavy gluon and the composite fermions needed to give mass to the 
bottom quark.

The lagrangian that describes our effective theory is the following (we work in the electroweak gauge-less limit, and omit the terms involving the 
$SU(2)_L \times U(1)_Y$ elementary gauge fields, which play no role in our analysis):
\begin{align}  \label{eq:Ltotal}
{\cal L} =&  \, {\cal L}_{elementary} + {\cal L}_{composite} + {\cal L}_{mixing} \\[0.7cm]
\label{eq:Lelem}
 {\cal L}_{elementary} =& \, -\frac{1}{4g_{el3}^2} \, G_{\mu\nu} G^{\mu\nu}  
                                    + \bar q^i_L i\!\Dslash\, q^i_L + \bar u^i_R i\!\Dslash\, u^i_R +  \bar d^i_R i\!\Dslash\, d^i_R   \\[0.5cm]
\label{eq:Lcomp}
\begin{split} 
 {\cal L}_{composite}  =&  
  \, -\frac{1}{4g_{*3}^2} \, G^*_{\mu\nu} G^{*\, \mu\nu}  
      + \frac{1}{2} \,\text{Tr}\left\{  \partial_\mu {\cal H}^\dagger \partial^\mu {\cal H} \right\} - V( {\cal H}^\dagger {\cal H}) \\[0.2cm]
  & + \text{Tr} \left\{ \bar{\cal Q} \left( i\!\!\dslash\, - \!\not\! G^*\! - {\bar m}_Q \right) {\cal Q}  \right\} + 
  \bar{\tilde T}  \left( i\!\!\dslash\, - \!\not \! G^*\! - {\bar m}_{\tilde T} \right) \tilde T   \\[0.2cm]
  & + \text{Tr} \left\{ \bar{\cal Q^\prime} \left( i\!\!\dslash\, - \!\not\! G^*\! - {\bar m}_{Q^\prime} \right) {\cal Q^\prime}  \right\} + 
  \bar{\tilde B}  \left( i\!\!\dslash\, - \!\not \! G^*\! - {\bar m}_{\tilde B} \right) \tilde B  \\[0.2cm]
  & - Y_{*U} \, \text{Tr} \{ \bar{{\cal Q}} \,  {\cal H} \} \, \tilde{T}  - Y_{*D} \, \text{Tr} \{ \bar{{\cal Q^\prime}} \,  {\cal H} \} \, \tilde{B} + h.c.
\end{split} \\[0.5cm]
\begin{split} \label{eq:Lmixing}
{\cal L}_{mixing} =&  
    \, \frac{1}{2}\, \frac{{\bar M}^2_{G_*}}{g_{*3}^2} \left( G_\mu - G_\mu^* \right)^2
    - \Delta_{L1}\, \bar q^3_L \left( T,B\right) - \Delta_{L2}\, \bar q^3_L \left( T^\prime,B^\prime\right)  \\[0.2cm]
   & - \Delta_{tR}\, \bar t_R \tilde T  - \Delta_{bR}\, \bar b_R \tilde B + h.c.  \, ,
\end{split}
\end{align}
where $V( {\cal H}^\dagger {\cal H})$ is the Higgs potential, and the derivative $D_\mu$ is covariant under $SU(3)_c$ transformations.
The superscript $i$ in eq.(\ref{eq:Lelem}) runs over the three SM families ($i =1,2,3$), with $q^3_L \equiv (t_L, b_L)$, $u^3_R \equiv t_R$, 
$d^3_R \equiv b_R$.
In addition to the Higgs mass and the self-couplings contained in the Higgs potential,
the lagrangian  (\ref{eq:Ltotal})  has thirteen free parameters: 
four composite masses in the fermionic sector ($\bar m_{Q}$, $\bar m_{Q^\prime}$, $\bar m_{\tilde T}$, $\bar m_{\tilde B}$); one for $G^*$ ($\bar M_{G^*}$);
four mass mixing terms ($\Delta_{L1}$, $\Delta_{L2}$, $\Delta_{tR}$, $\Delta_{bR}$); two composite Yukawa couplings ($Y_{*U}$, $Y_{*D}$); and
two gauge couplings ($g_{el3}$, $g_{*3}$).

By construction, the elementary fields couple to the composite ones only through the mass mixing lagrangian ${\cal L}_{mixing}$.
This implies that the SM Yukawa couplings arise only through the couplings of the Higgs to the composite fermions, and their
mixings to the elementary fermions. 
We consider the case in which the strong sector is flavor anarchic~\cite{Grossman:1999ra,Gherghetta:2000qt,Huber:2000ie}.
Under this assumption
the hierarchy in the  masses and mixings of the SM quarks follows from the hierarchy in the composite/elementary mixing 
parameters.
Reproducing the full spectrum of quark masses and the Cabibbo-Kobayashi-Maskawa matrix requires introducing 
in the lagrangian three families of heavy fermions and three sets of mixing terms with coefficients 
$\{\Delta^i_{L1}$, $\Delta^i_{L2}$, $\Delta^i_{uR}$, $\Delta^i_{dR}\}$, one for each SM flavor $i$.
In this case, on the other hand, the mixing parameters of the light elementary quarks are small and 
their effect can be neglected  in studying the $G^*$ phenomenology,
so that one can focus on just the third generation of
composite fermions.~\footnote{In fact, once produced, heavy fermions of the first two generations will also decay mostly to tops and bottoms,
since flavor-changing transitions are not suppressed in the strong sector, while the couplings to the light SM quarks are extremely small,
see the discussion in Ref.~\cite{Contino:2006nn}. In this sense, the addition of the first two generations of heavy fermions 
in the lagrangian~(\ref{eq:Ltotal}) would not qualitatively
alter the phenomenology described in this paper.} These two simplifications have in fact been applied   
in writing eqs.(\ref{eq:Lcomp}),(\ref{eq:Lmixing}),  and are 
justified a posteriori by our assumption of flavor anarchy in the strong sector.~\footnote{If the strong sector is not anarchic, the phenomenology
can change significantly and additional signatures, like for example the unsuppressed decay of $G^*$ to light quarks, can appear, see for example
Ref.~\cite{Redi:2011zi}.
Our analysis, on the other hand, can still be relevant although the decay modes considered here might not be the dominant ones.}
The `anarchic' scenario  has been extensively studied in the framework of 5D warped models, 
see Refs.~\cite{Huber:2003tu,Agashe:2004cp,Csaki:2008zd,Blanke:2008zb,Agashe:2008uz,Gedalia:2009ws}.  
Although flavor-changing neutral effects are parametrically suppressed by the degree of compositeness of the SM fermions,
a mechanism which has been dubbed RS-GIM flavor protection~\cite{Agashe:2004cp}, it has been shown that 
numerically strong bounds on the masses of the KK gluon and of the KK fermions still generically follow from 
CP-violating observables in the kaon sector, namely $\epsilon_K$~\cite{Csaki:2008zd,Blanke:2008zb} and 
$\epsilon^\prime/\epsilon$~\cite{Gedalia:2009ws}, the neutron electric dipole moment~\cite{Agashe:2004cp}, and $b\to s\gamma$~\cite{Agashe:2008uz}.
Several solutions have been proposed to evade these bounds, without renouncing to the explanation of the SM flavor hierarchy,
in which the strong sector (\textit{i.e.} the bulk of 5D models) is assumed to be invariant under additional flavor 
symmetries~\cite{Fitzpatrick:2007sa,Santiago:2008vq,Csaki:2008eh,Csaki:2009wc}
or to preserve CP~\cite{Redi:2011zi}. In these scenarios the KK gluon can have a mass as light as a few TeVs and its phenomenology is not qualitatively
modified. In the following we will assume that some mechanism is at work to alleviate the flavor bounds and we will 
estimate the LHC reach on the heavy masses without imposing any restriction on them.

We assume $\Delta_{L2} \ll  \Delta_{L1}$, which can  naturally follow, for example, from the RG flow in the full theory~\cite{Contino:2006qr}.
In this limit, up to small $O(\Delta_{L2})$ effects, the $P_{LR}$ parity of the strong sector is not broken by the coupling of the elementary $b_L$ 
to the latter, and the vertex $Zb\bar b$ is protected by large tree-level corrections~\cite{Agashe:2006at}.~\footnote{It has been also pointed out 
in Ref.~\cite{Mrazek:2011iu} that in the case of $SO(5)/SO(4)$ theories the $P_{LR}$ parity is accidental  at low energy if the fermions transform as 
fundamental representations of $SO(5)$.  This ensures automatic protection to $Zb\bar b$ even if the strong dynamics is not $P_{LR}$ invariant at 
the fundamental level.}
In fact, this protection was the main motivation to require the strong sector to be invariant under $P_{LR}$ and choose 
the fermionic representations of eq.(\ref{eq:fermions}).
We further require $\Delta_{tR} \sim \Delta_{bR}$, so that the small  ratio $m_b/m_t$ follows  from the hierarchy $\Delta_{L2}/\Delta_{L1} \ll 1$. 
Notice also that for $\Delta_{L2} \to 0$ the two-site model of Ref.~\cite{Contino:2008hi} is contained as a subsector of the TS5 model.

In order to diagonalize the two-site lagrangian (\ref{eq:Ltotal}) one has to make a field rotation
from the elementary/composite basis to the mass eigenstate basis~\cite{Contino:2006nn}. The only exception is given by 
the first two generations of elementary quarks, since they do not mix with the composite fermions and can thus be directly identified
with the corresponding SM states.
We performed the field rotation analytically before EWSB by treating $\Delta_{L2}$ as a small parameter and including only the leading order
terms in the perturbative expansion. The final lagrangian is shown in eqs.(\ref{eq:Lrotated})-(\ref{eq:LrotatedHiggs}) of Appendix~\ref{app:Lagrangian}. 
We use an economical notation 
where symbols denoting elementary (composite) fields before the rotation now indicate the SM (heavy) fields.
More details will be reported in a separate publication~\cite{wp}. 
Here we  just  list the relevant parameters that control the spectrum and the interactions of the
mass eigenstates and discuss the numerical assumptions we made on them.
The elementary/composite rotation in the fermionic sector can be parametrized in terms of six mixing angles, 
which can be conveniently defined as follows:
\begin{equation}
\tan\vphi_{tR} = \frac{\Delta_{tR}}{\bar m_{\tilde T}}\, , \qquad
\tan\vphi_{bR} = \frac{\Delta_{bR}}{\bar m_{\tilde B}}\, , \qquad
\tan\vphi_{L} = \frac{\Delta_{L1}}{\bar m_{Q}}\, , \qquad
s_2 = \frac{\Delta_{L2}}{\bar m_{Q^\prime}} \cos\vphi_L\, , 
\end{equation}
plus the two mixing parameters defined in eq.(\ref{eq:s3s4}) of Appendix~\ref{app:Lagrangian}.
Here $\sin\vphi_{tR}$, $\sin\vphi_{bR}$, $\sin\vphi_{L}$ respectively denote the degree of compositeness of  
$t_R$, $b_R$ and $(t_L, b_L)$. In the gauge sector,  the elementary and the composite couplings 
$g_{el3}$, $g_{*3}$ determine the rotation angle $\theta_3$ and the SM gauge coupling $g_3$ as follows:
\begin{equation}
\label{eq:vecrot}
\tan\theta_3 = \frac{g_{el3}}{g_{*3}}\, , \qquad g_3 = g_{el3} \cos\theta_3 = g_{*3} \sin\theta_3 \, .
\end{equation}
The physical masses are given by
\begin{gather}
\label{eq:masses1}
m_{\tilde T} = \frac{\bar m_{\tilde T}}{\cos\vphi_{tR}}\, , \qquad 
m_{\tilde B} = \frac{\bar m_{\tilde B}}{\cos\vphi_{bR}}\, , \qquad 
m_{T} = m_{B} = \frac{\bar m_{Q}}{\cos\vphi_{L}}\, , \qquad 
m_{T^\prime} = m_{B^\prime} \simeq \bar m_{Q^\prime} \, , \\[0.35cm]
\label{eq:masses2}
m_{T_{5/3}} = m_{T_{2/3}} = m_T \cos\vphi_L \, , \qquad m_{B_{-4/3}} = m_{B_{-1/3}} = m_{T^\prime}\, ,
\end{gather}
where only the first four are independent, and
\begin{equation}
M_{G^*} = \frac{{\bar M}_{G^*}}{\cos\theta_3}\, .
\end{equation}
There are thus thirteen independent angles and physical masses (in addition to the Higgs mass and self-couplings), 
corresponding to the initial number of  parameters of the lagrangian~(\ref{eq:Ltotal}).

After the EWSB, the SM top and bottom quarks acquire a mass, and the heavy masses in eqs.(\ref{eq:masses1}), (\ref{eq:masses2}) get corrections 
of order $(vY_*/{\bar m})^2$, where $v = 246\,$GeV is the electroweak scale. In the following, we assume $r \equiv (vY_*/{\bar m}) \ll 1$ and 
compute all quantities at leading order in $r$.~\footnote{\label{ftn:rexp}
More specifically, we compute the SM top and bottom masses and all decay widths of the heavy fermions at $O(r)$,  and we neglect 
the $O(r^2)$ EWSB corrections to the masses of the heavy fermions.} This is in fact a required condition in order to consistently neglect 
operators with a higher number of Higgs fields in the two-site effective lagrangian. 
We thus work under the hypothesis of parametrically small $r$, although in some 
of the benchmark points that we will consider $r$ is numerically not  very small.~\footnote{In the most 
extreme case under consideration $Y_* =3$, $\bar m = 1\,$TeV, so that $r\simeq 3/4$.}
In those cases we  expect that our phenomenological analysis still gives an accurate description at the qualitative level,
although our formulas can receive quantitatively important corrections.
At leading order in $r$, the masses of the SM top and bottom quarks read
\begin{equation}
m_t = \frac{v}{\sqrt{2}} Y_{*U} \sin\varphi_L \sin\varphi_{tR}\, , \qquad  
m_b = \frac{v}{\sqrt{2}} Y_{*D} s_2 \sin\varphi_{bR}\, .
\end{equation}

Fixing $m_t$, $m_b$ and $g_3$ to their experimental values gives three conditions on the set of parameters.
As a further simplification, in what follows we will set
\begin{equation}
Y_{*U} = Y_{*D} = Y_* \, , \qquad
\sin\varphi_{bR} = \sin\varphi_L \, , \qquad
m_{T^\prime} = m_T = m_{\tilde B} = m_{\tilde T} \, ,
\end{equation}
so that there are five free parameters left: three mixing angles ($\sin\varphi_{L}$, $\sin\varphi_{tR}$, $\tan\theta_3$) and two masses 
($m_{\tilde T}$, $M_{G^*}$). Alternatively,  one can also trade  two of the mixing angles, $\sin\varphi_{L}$, $\tan\theta_3$, for
the two couplings $Y_*$, $g_{*3}$.

\subsection{Phenomenology of $G^*$}
\label{subsec:Gstarpheno}

\begin{figure}[tbp]
\begin{center}
\includegraphics[width=0.6\textwidth,clip,angle=0]{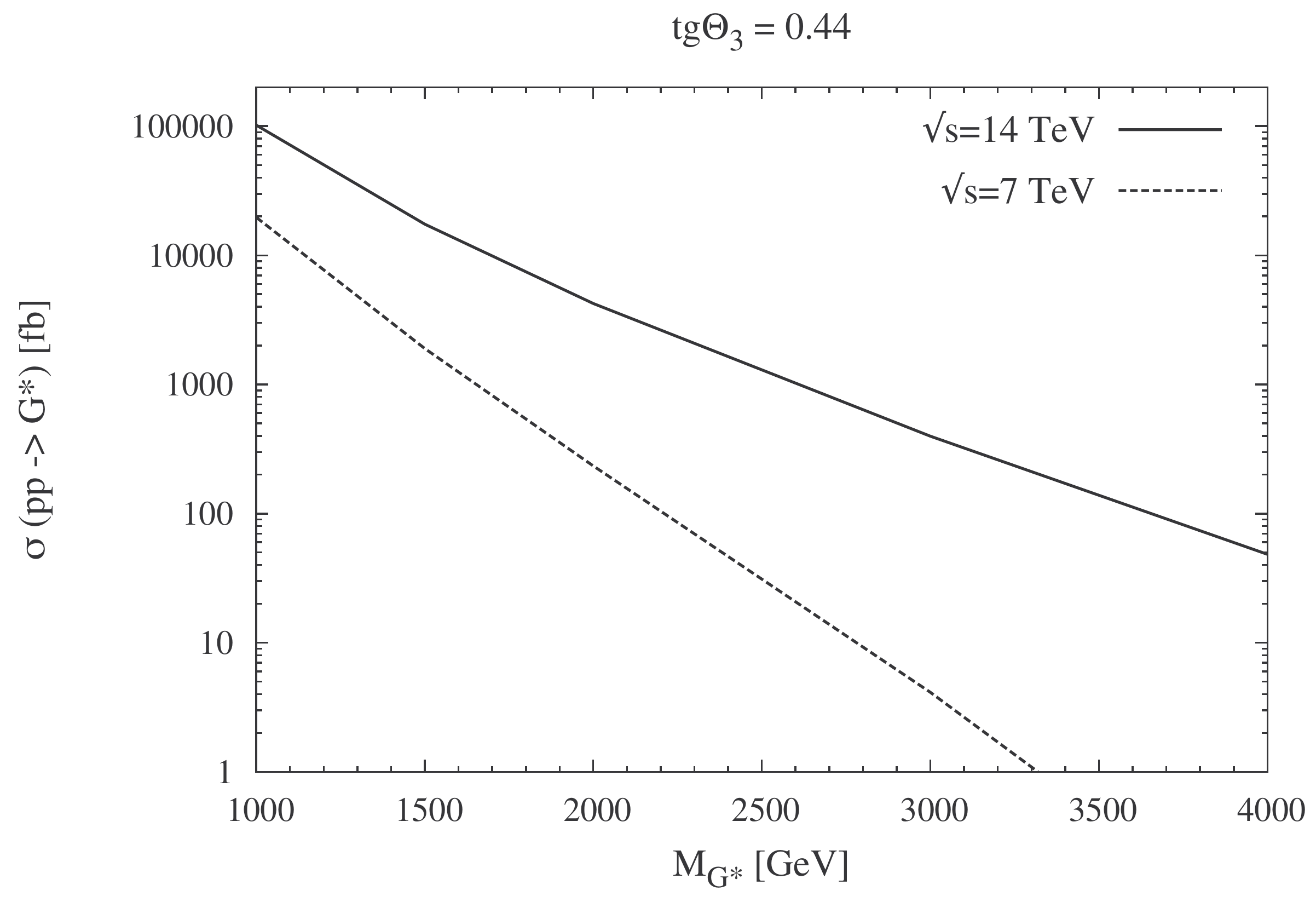}
\caption[]{
\label{fig:Gstarxsec}
\small
Cross section of the  Drell-Yan production of $G^*$ at the LHC, $pp \to q\bar q \to G^*$,  for $\tan\theta_3 = 0.44$ ($g_{*3}=3$). 
For different values of $\tan\theta_3$ the cross section scales as $(\tan\theta_3)^2$.
}
\end{center}
\end{figure}

\begin{figure}[tbp]
\begin{center}
\includegraphics[width=0.482\textwidth,clip,angle=0]{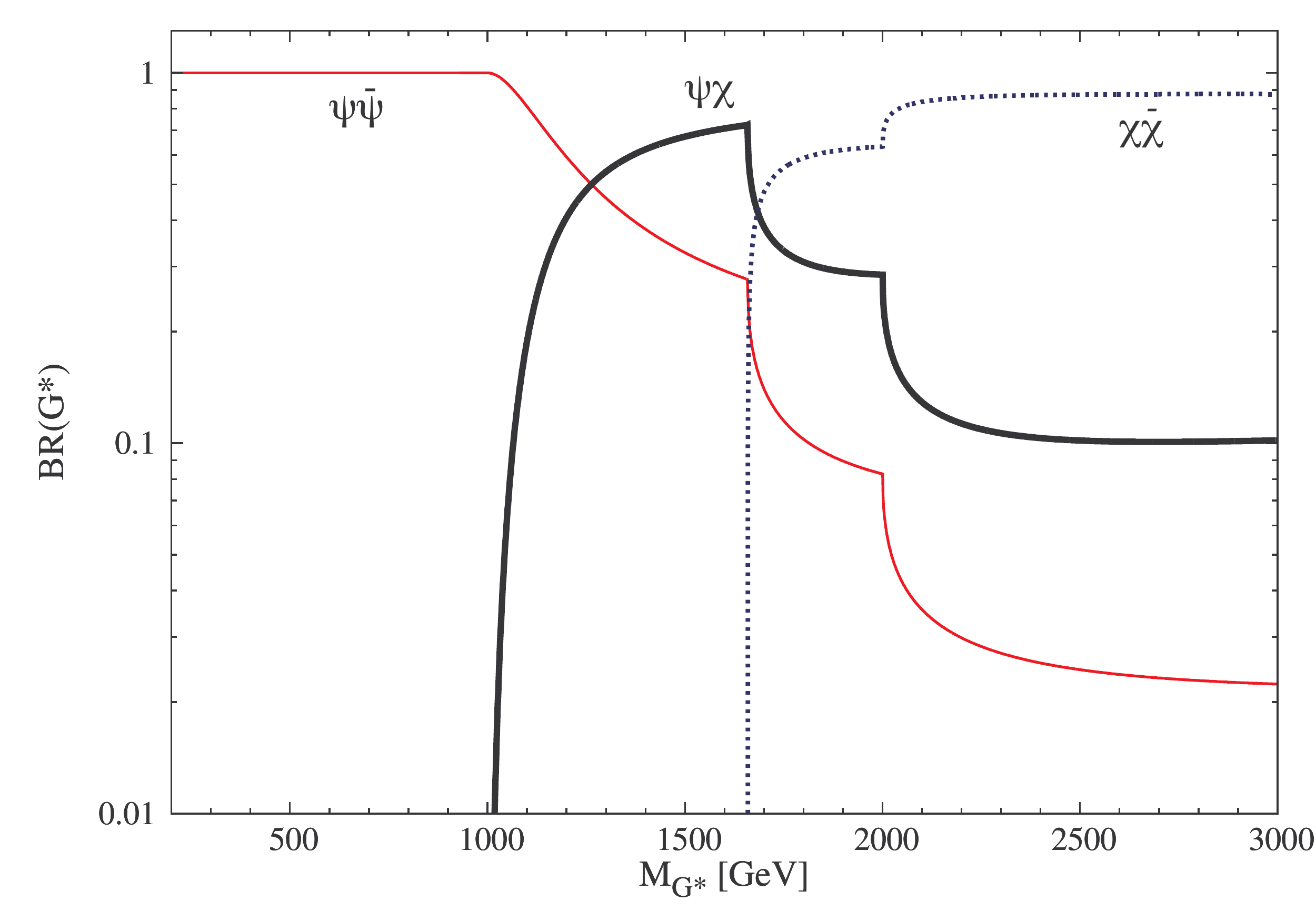}
\includegraphics[width=0.505\textwidth,clip,angle=0]{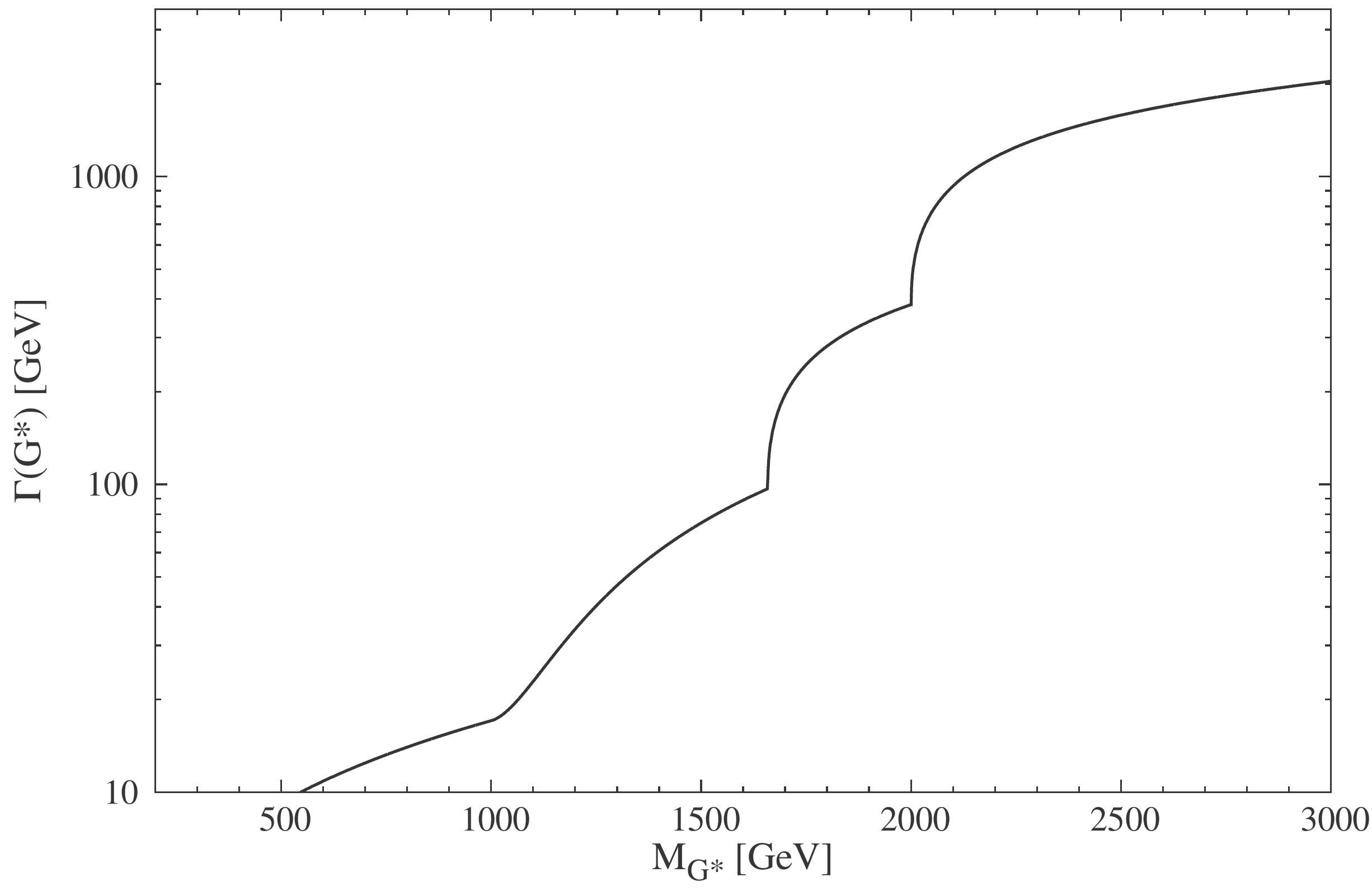}
\caption[]{
\label{fig:GstarDecay}
\small
Left plot: branching ratios for the decay of $G^*$ to two SM fermions, $\psi\bar\psi$, two heavy fermions, $\chi\bar\chi$,
and one SM plus one heavy fermion, $\psi\bar\chi+\chi\bar\psi$, as a function of $M_{G^*}$.
Right plot: total decay width of $G^*$ as a function of $M_{G^*}$. Both plots are done setting the other parameters to the following
reference values: $m_{\tilde T} =1\,$TeV and $\tan\theta_3 = 0.44$, $\sin\varphi_{tR} = 0.6$, $Y_* =3$.
}
\end{center}
\end{figure}

Starting from the  rotated lagrangian~(\ref{eq:Lrotated}), one can derive the phenomenology of the heavy gluon.
The couplings of $G^*$ to the fermions are reported  in eqs.(\ref{eq:Gqq})-(\ref{eq:Ghh}) of  Appendix~\ref{app:Lagrangian}.
In particular, $G^*$ couples to the light quarks with strength $(-g_3 \tan\theta_3)$, and in general with a much stronger coupling to the top, 
bottom and heavy fermions. Gauge invariance forbids a coupling with two gluons and one $G^*$ at the level of renormalizable operators.
The main rate of production at the LHC hence comes from the Drell-Yan process $pp \to q\bar q \to G^*$. 
Figure~\ref{fig:Gstarxsec}  shows the relative cross section for a reference value $\tan\theta_3=0.44$  (corresponding to $g_{*3} =3$).
The analytic expression of the $G^*$ decay partial widths is given in eqs.(\ref{eq:GammaGqq})-(\ref{eq:GammaGhh}) of Appendix~\ref{app:Lagrangian}.
%
%%
%%%%%%%%%%%%%%%%%%%%%%%%%%%%%%%%%%%%%%%%%%%%%%%%%%%%%%%%%%%%%%%%%%%%%%%
%%%%%%%%%%%%%%%%%%%%%%%%%%%%%%%%%%%%%%%%%%%%%%%%%%%%%%%%%%%%%%%%%%%%%%%
%%
When kinematically allowed,  $G^*$ can decay to two SM fermions, $\psi\bar\psi = q\bar q + t\bar t+b\bar b$, where $q =u,d,c,s$;
two heavy fermions, $\chi\bar\chi$, where $\chi$ denotes any of the heavy fermions;
or one heavy plus one SM fermion, $\psi\bar\chi + \chi\bar\psi = T\bar t_L + B\bar b_L + \tilde T \bar t_R + \tilde B \bar b_R +c.c.$.~\footnote{We
classify all particles according to their $SU(2)_L \times U(1)_Y$ quantum numbers, and compute all decay widths at $O(r)$.  See also
footnote~\ref{ftn:rexp}.}
The plot on the left of Fig.~\ref{fig:GstarDecay} shows
the relative branching ratios as functions of $M_{G^*}$ for reference values $m_{\tilde T} =1\,$TeV and 
$\tan\theta_3 = 0.44$, $\sin\varphi_{tR} = 0.6$, $Y_* =3$ (hence $\sin\varphi_{L} = 0.56$).
The corresponding  total decay width of $G^*$ is shown in the plot on the right of the same figure, for the same reference values of parameters.
%%
%%%%%%%%%%%%%%%%%%%%%%%%%%%%%%%%%%%%%%%%%%%%%%%%%%%%%%%%%%%%%%%%%%%%%%%
%%%%%%%%%%%%%%%%%%%%%%%%%%%%%%%%%%%%%%%%%%%%%%%%%%%%%%%%%%%%%%%%%%%%%%%

When the decays to one heavy fermion are kinematically forbidden, $M_{G^*} \lesssim m_{\tilde T}$,  the heavy gluon can decay only to pairs
of SM quarks. The relative importance of the various channels is controlled by $\tan\theta_3$ and the top degrees of compositeness
$\sin\varphi_{tR}$, $\sin\varphi_{L}$. For small values of $\tan\theta_3$,
which are naturally implied by the hierarchy of couplings $g_{el3} \ll g_{*3}$, or large top degree of compositeness,
the dominant channel is $t\bar t$. 
For example, for a fully composite $t_R$, $\sin\varphi_{tR} =1$, and $\tan\theta_3 =0.2$, $Y_* =3$ (hence
$\sin\varphi_{L} = 0.33$) one has $BR(G^* \to t\bar t) = 0.98$, $BR(G^* \to q\bar q) = 0.012$.
In this limit, $t\bar t$ is the best channel to discover the $G^*$, despite the huge QCD background~\cite{Agashe:2006hk,Lillie:2007yh}.
On the other hand, large branching ratios to pairs of light quarks can be  obtained even for moderate top degrees of compositeness
if $\tan\theta_3$ is not too small.
This is a consequence of the cancellation that can occur between the two terms in eq.(\ref{eq:Gpsipsi}) due to the relative minus sign.
For example, for $\sin\varphi_{tR} =0.6$, $\tan\theta_3 =0.44$, $Y_* =3$  one has $BR(G^* \to t\bar t) = 0.18$, $BR(G^* \to q\bar q) = 0.69$.
In this case the strongest discovery reach (or exclusion power) comes from the dijet searches.

For large values of $M_{G^*}$, the heavy gluon mostly decays to pairs of heavy fermions, due to the large number of available
channels and the large couplings. This is clearly illustrated by the plot on the left of Fig.~\ref{fig:GstarDecay}.
In this case a first threshold to $T_{5/3}\bar T_{5/3} + T_{2/3}\bar T_{2/3}$ opens up at $M_{G^*} = 2 m_{T_{5/3}} = 2 m_{\tilde T} \cos\varphi_L \simeq 1.66\,$TeV,
and a second one at $M_{G^*} = 2 m_{\tilde T} = 2\,$TeV. In this limit the large value of the total decay width that follows for the large
multiplicity of decay channels (see the right plot of Fig.~\ref{fig:GstarDecay}) makes a discovery of $G^*$ very challenging.

\begin{figure}[tbp]
\begin{center}
\includegraphics[width=0.6\textwidth,clip,angle=0]{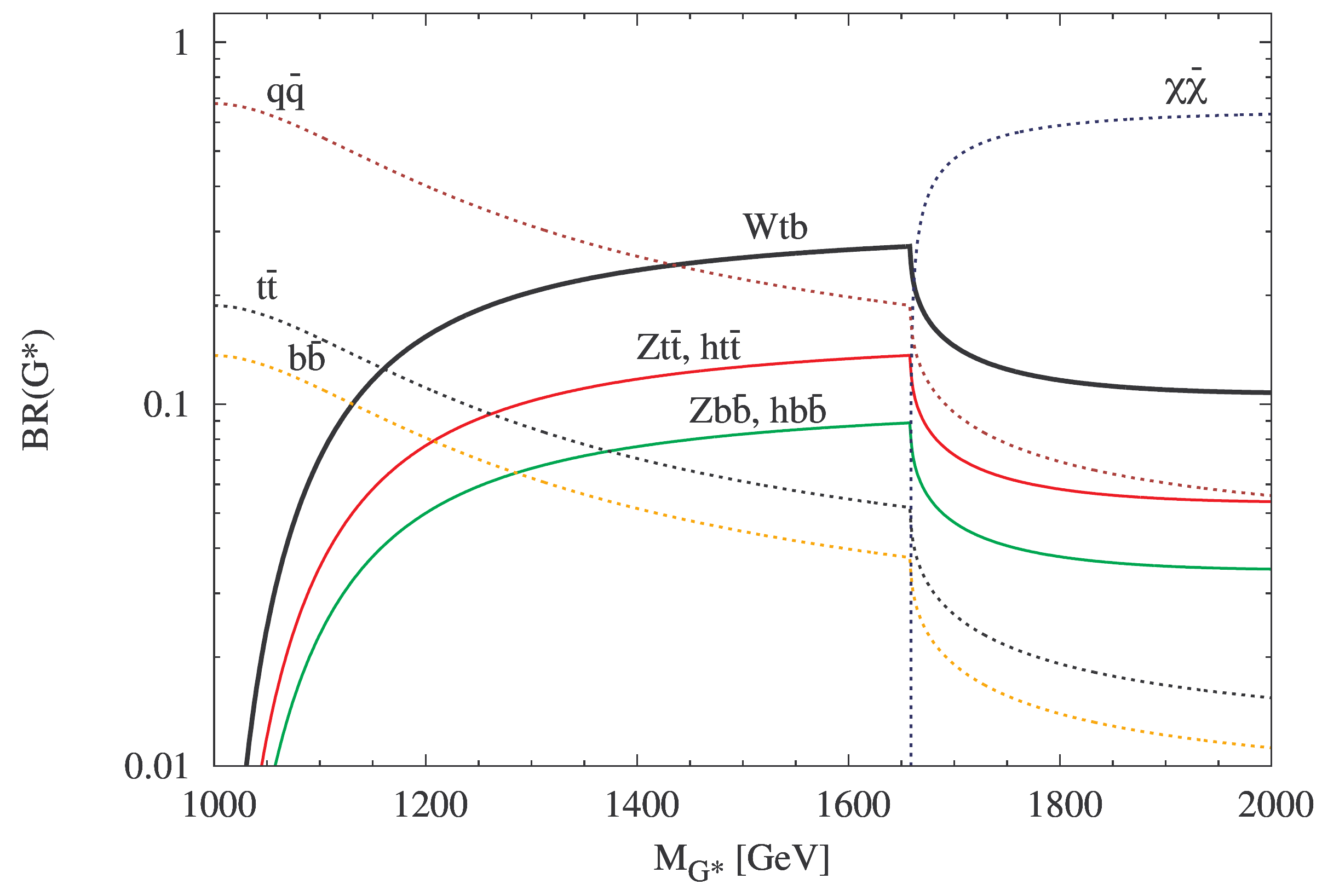}
\caption[]{
\label{fig:hldecays}
\small
Branching ratio of the various final states that follow from the decay of $G^*$, as functions of $M_{G^*}$.
The other parameters are set to the same reference values used for Fig.~\ref{fig:GstarDecay}:
$m_{\tilde T} =1\,$TeV and $\tan\theta_3 = 0.44$, $\sin\varphi_{tR} = 0.6$, $Y_* =3$.
The decay channel $ht\bar t$ ($hb\bar b$) has the same branching ratio of $Zt\bar t$ ($Zb\bar b$).
}
\end{center}
\end{figure}

There is however a third scenario, which is realized for $m_{\tilde T} \lesssim M_{G^*} < 2 m_{T_{5/3}}$. In this case the threshold for
decaying to two heavy fermions is kinematically closed, so that the $G^*$ total width remains small, but decays
to one SM fermion plus one heavy fermion are allowed and have a large branching ratio, see Fig.~\ref{fig:GstarDecay}.
Such heavy-light decays are possible only if the SM fermion and the heavy fermion have the same quantum numbers under
$SU(2)_L \times U(1)_Y$, that is, the heavy fermion must be one of the partners of the top or bottom: $\chi = T, B, \tilde T, \tilde B$.
Due to their large Yukawa coupling, these heavy states mostly decay to one top or bottom plus one longitudinally-polarized
vector boson or  Higgs boson.
The analytic expression of the corresponding decay widths is reported in eqs.(\ref{eq:GHW})-(\ref{eq:GHh}) of Appendix~\ref{app:Lagrangian}.
In the limit $m_\chi \gg m_{t}, m_W$, the branching ratios of the possible decay modes are:
\begin{equation}
\label{eq:fermionBRs}
\begin{gathered}
 BR\left(T\rightarrow Z_L t_R\right)\simeq  BR\left(T\rightarrow h t_R\right)\simeq 50\% \, , \\[0.25cm]
 BR\left(B\rightarrow W_Lt_R\right)\simeq 100\% \, , \\[0.25cm]
 BR\big(\tilde{T}\rightarrow W_Lb_L\big)\simeq 50\% \, , \quad  
 BR\big(\tilde{T}\rightarrow Z_Lt_L\big)\simeq  BR\big(\tilde{T}\rightarrow ht_L\big)\simeq 25\% \, , \\[0.25cm]
 BR\big(\tilde{B}\rightarrow Z_Lb_R\big)\simeq BR\big(\tilde{B}\rightarrow hb_R\big)\simeq 50\% \, .
\end{gathered}
\end{equation}
Considering the pattern of  decays, we can identify five final channels:
\begin{equation} \label{eq:finalch}
Bb, \tilde Tt  \to Wtb  \, , \qquad 
T t , \tilde T  t\to Z t\bar t \, , \qquad 
T t , \tilde T  t\to h t\bar t \, , \qquad 
\tilde B b      \to Z b\bar b \, , \qquad 
\tilde B b      \to h b\bar b \, .
\end{equation}
Their relative importance is shown in Fig.~\ref{fig:hldecays} as a function of $M_{G^*}$, for the same reference values of parameters
chosen for Fig.~\ref{fig:GstarDecay}.
%%
%%%%%%%%%%%%%%%%%%%%%%%%%%%%%%%%%%%%%%%%%%%%%%%%%%%%%%%%%%%%%%%%%%%%%%%
%%%%%%%%%%%%%%%%%%%%%%%%%%%%%%%%%%%%%%%%%%%%%%%%%%%%%%%%%%%%%%%%%%%%%%%
%%
One can see that each of the heavy-light topologies has a sizable branching ratio in a large range of $G^*$ masses above 
$m_{\tilde T}$ and below $2m_{T_{5/3}}$, while the $t\bar t$ channel is suppressed compared to its value at $M_{G^*} < m_{\tilde T}$. 
Notice that although $BR(G^* \to q\bar q)$ gets suppressed as well,
it is still sizable for the benchmark values of parameters chosen in Fig.~\ref{fig:hldecays}. 
While large branching fractions for the heavy-light decays are mainly implied by the kinematics and as such are  a robust prediction,
the value of $BR(G^* \to q\bar q)$ is strongly dependent on $\tan\theta_3$ and can thus be easily made small.
%%
%%%%%%%%%%%%%%%%%%%%%%%%%%%%%%%%%%%%%%%%%%%%%%%%%%%%%%%%%%%%%%%%%%%%%%%
\begin{figure}[tbp]
\begin{center}
\includegraphics[width=0.495\textwidth,clip,angle=0]{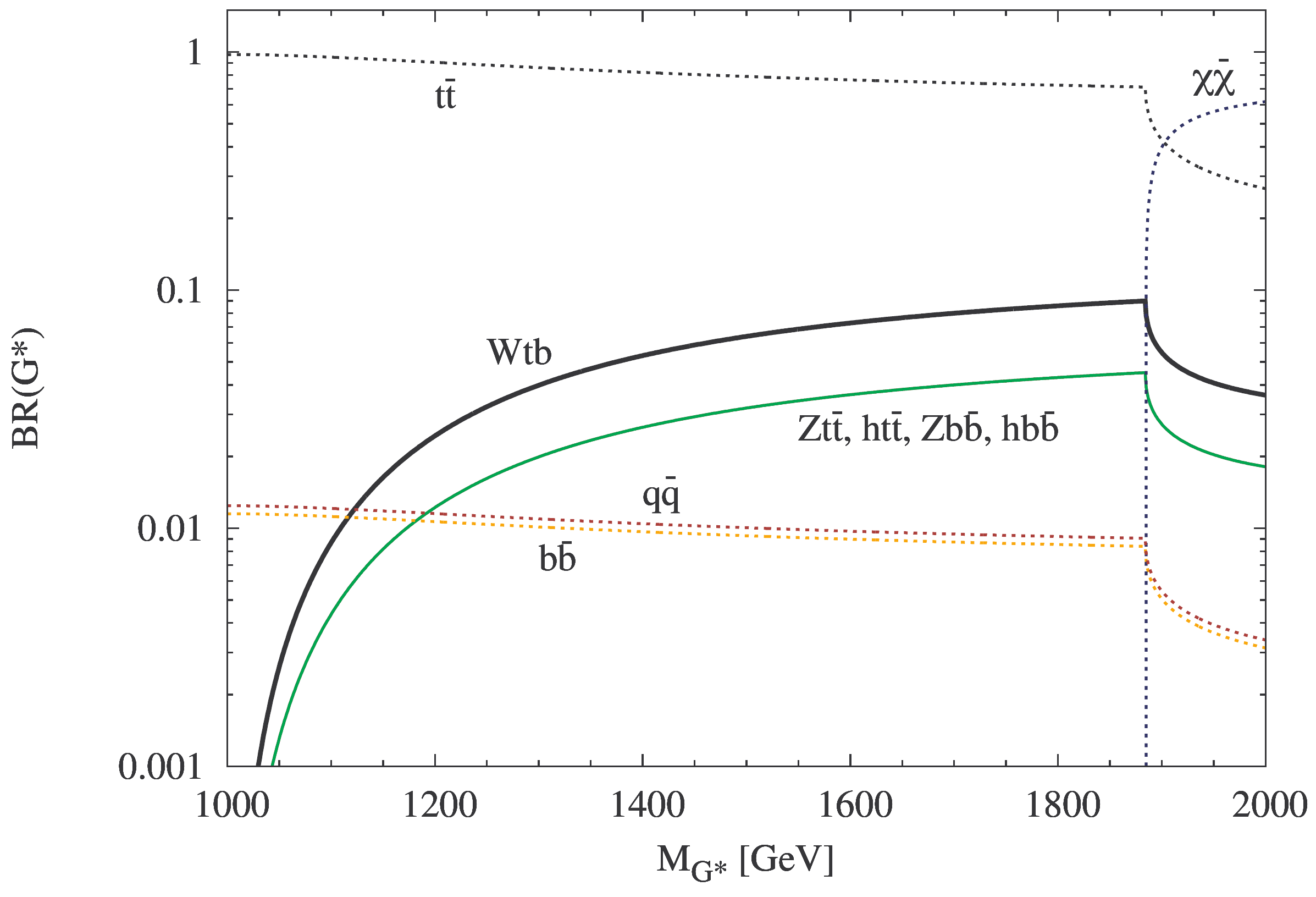}
\includegraphics[width=0.495\textwidth,clip,angle=0]{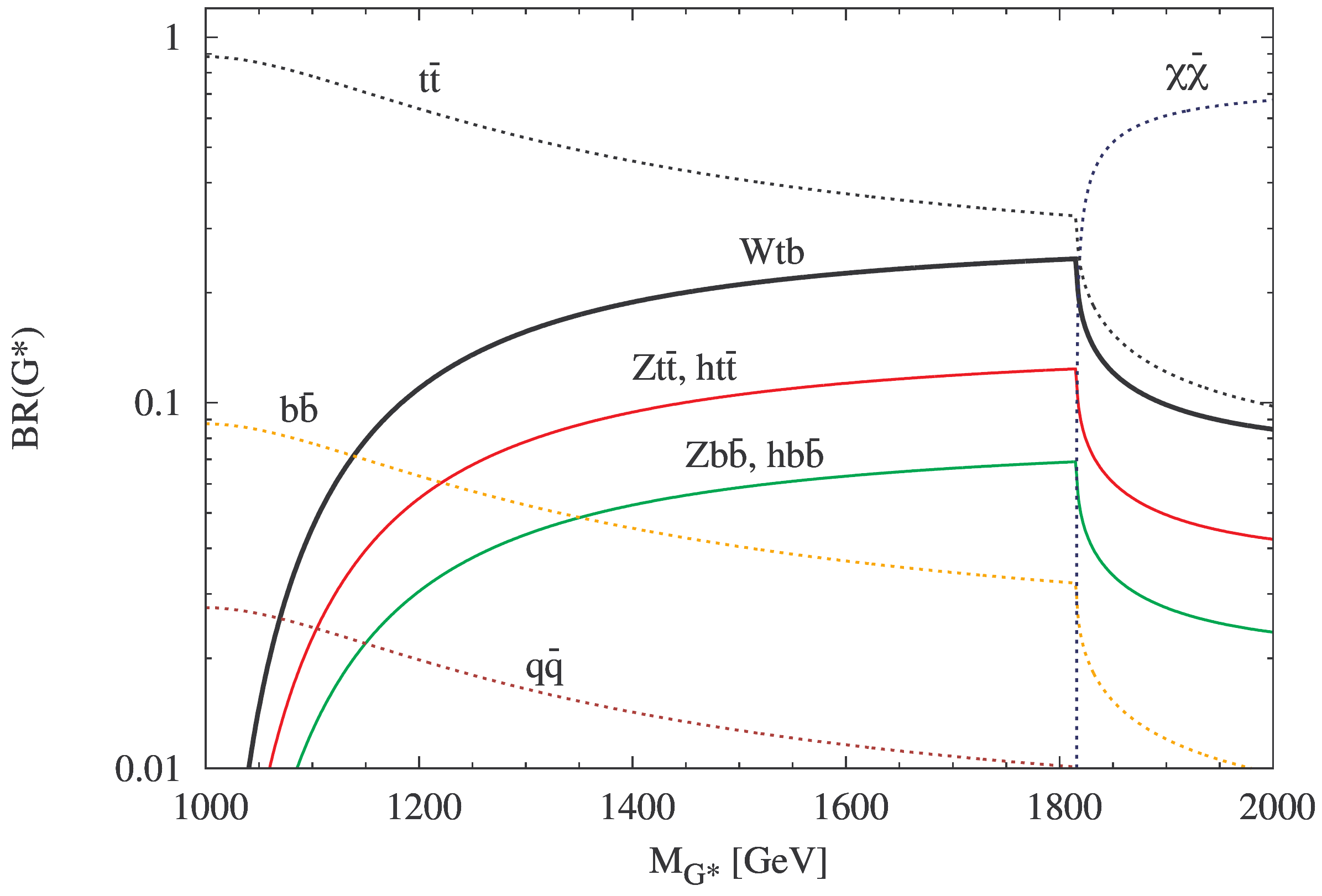}
\caption[]{
\label{fig:hldecays2}
\small
Branching ratios of $G^*$ to the various final states as functions of $M_{G^*}$.
Compared to Fig.~\ref{fig:hldecays}, two different set of parameters have been chosen:  
$\sin\varphi_{tR} = 1$, $\tan\theta_3 =0.2$, $Y_* =3$, $\sin\varphi_{L} =0.33$ 
(left plot), and $\sin\varphi_{tR} = 0.8$, $\tan\theta_3 =0.2$, $Y_* =3$, $\sin\varphi_{L} =0.41$  (right plot).
Both plots are done fixing $m_{\tilde T} =1\,$TeV.
}
\end{center}
\end{figure}
%%%%%%%%%%%%%%%%%%%%%%%%%%%%%%%%%%%%%%%%%%%%%%%%%%%%%%%%%%%%%%%%%%%%%%%
%%
This is clearly illustrated in Fig.~\ref{fig:hldecays2}, where the branching ratios to the various channels are shown for two different choices
of parameters. The left plot refers to the same benchmark point adopted in Ref.~\cite{Agashe:2006hk},  $\sin\varphi_{tR} = 1$, $\tan\theta_3 =0.2$, 
$Y_* =3$ ($\sin\varphi_{L} = 0.33$); in this case the $t\bar t$ channel largely dominates over the others until the threshold 
$M_{G^*} = 2 m_{T_{5/3}}$, as a consequence of the full degree of compositeness of the right-handed top.
The right plot refers instead to a similar set of parameters where, however, the degree of compositeness of $t_R$ is slightly smaller:
$\sin\varphi_{tR} = 0.8$, $\tan\theta_3 =0.2$, $Y_* =3$ ($\sin\varphi_{L} = 0.41$). It is evident how in this case the $t\bar t$ branching ratio
is substantially reduced in the range $m_{\tilde T} \lesssim M_{G^*} < 2 m_{T_{5/3}}$, and those of the heavy-light channels, especially $Wtb$, are sizable.
Notice that in this case the branching ratio of $q\bar q$ is extremely small, as due to the small value of $\tan\theta_3$.

Given their numerical importance for $m_{\tilde T} \lesssim M_{G^*} < 2 m_{T_{5/3}}$, the heavy-light decay channels represent a new promising
strategy for discovering the $G^*$, in a limit in which the latter is still a relatively narrow resonance. 
The reason is mainly  twofold: on the one hand, one can use the peculiar topology of the heavy-light decays, where the invariant
mass of a subsystem reconstructs the physical mass of the heavy fermion, to efficiently reduce the QCD background; on the other hand,
the $t\bar t$ channel has a much smaller rate for $m_{\tilde T} \lesssim M_{G^*} < 2 m_{T_{5/3}}$, hence it is much less efficient as
a discovery channel.
Quite remarkably, as it will emerge from our analysis,
the heavy-light decays seem  extremely competitive  channels for discovering the heavy fermions as well. As a final bonus, 
they also give the opportunity to measure some important features of the fermionic sector,
like for example the couplings of the heavy fermions to the SM vector bosons, and thus determine its origin.

Ideally, one could look for the presence of (fermionic) resonances in all the  channels of eq.(\ref{eq:finalch}).
Simple considerations on the size of the SM backgrounds suggest that the most promising final states could be the following
(we assume that the Higgs boson is light and mostly decays to $b\bar b$): $Wtb$;
$Zt\bar t, h t\bar t \to b\bar b t\bar t$;  $Zb\bar b, h b\bar b \to 4b$.
A heavy fermion with charge $+2/3$ will thus appear as a $(b\bar bt)$ resonance in $b\bar bt\bar t$ events, and 
possibly as a $(Wb)$ resonance in $Wtb$ events if it is a singlet of $SU(2)_L$ ($\tilde T$). The non-observation of this latter
signal would in turn point in favor of a classification of the heavy top as part of a doublet of $SU(2)_L$ ($T$).
A heavy fermion with charge $-1/3$, on the other hand, will appear as a $(3b)$ resonance in $4b$ events
only if it is an electroweak singlet ($\tilde B$) \textit{and} the SM $b_R$ is not too elementary (this was our assumption
when we derived the estimates in eq.(\ref{eq:fermionBRs})). The $(4b)$ channel
can thus give a way to test the degree of compositeness of $b_R$. If $b_R$ is very much elementary, on the other hand,
even a singlet $\tilde B$ will decay mostly to $Wtb$ events, similarly to a doublet $B$. In this case one will observe a $(Wt)$
resonance in $Wtb$ events.

The above discussion shows how a detailed picture on the quantum numbers of the fermionic resonances can be obtained by
exploiting all the decay final states. In what follows we will consider the $Wtb$ channel, which seems to be the most promising one,
and we will study it in detail.

%%%%%%%%%%%%%%%%%%%%%%%%%%%%%%%%%%%%%%
\section{Analysis of the $Wtb$ channel at the LHC}
\label{sec:analysis}
%%%%%%%%%%%%%%%%%%%%%%%%%%%%%%%%%%%%%%

In this section we discuss the prospects of observing the signal in the $Wtb$ channel at the LHC.
We will present a simple parton-level analysis aimed at assessing the LHC discovery reach, and focus on  final states with one lepton.
Our event selection will be cut-based and driven by simplicity as much as possible. We consider two center-of-mass energies: $\sqrt{s}=7\,$TeV,
the energy of the current phase of data taking, and $\sqrt{s}=14\,$TeV,  the design energy that will
be reached in the second phase of operation of the LHC.  Our selection strategy and the set of kinematic cuts that we will design 
are largely independent, however, of the value of the collider energy.  This is because they will be  optimized to exploit the peculiar kinematics 
of the signal, and a change in the collider energy mainly implies a rescaling of the  production cross sections  of signal and background
via the parton luminosities, without affecting the kinematic distributions.

\begin{figure}[tbp]
\begin{center}
\includegraphics[width=0.44\textwidth,clip,angle=0]{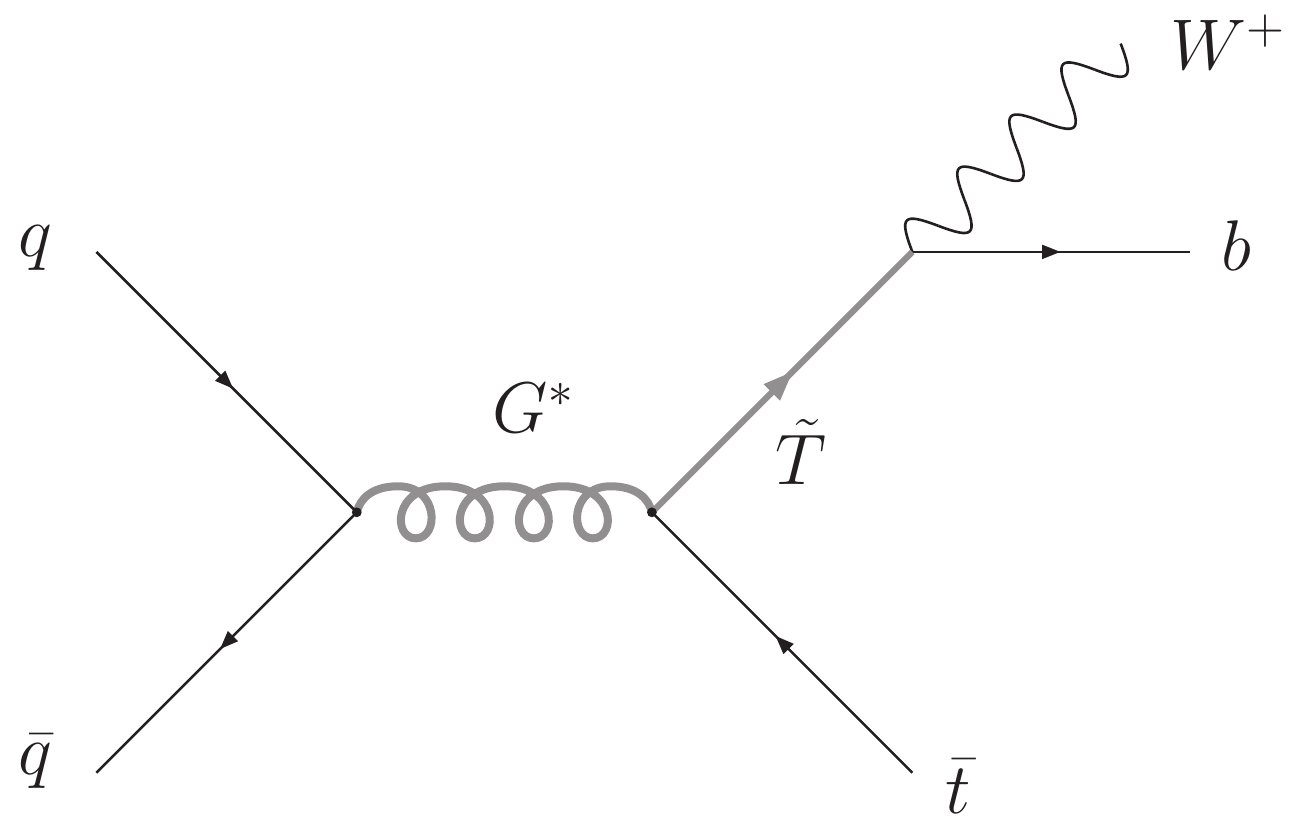}
\hspace{0.85cm}
\includegraphics[width=0.44\textwidth,clip,angle=0]{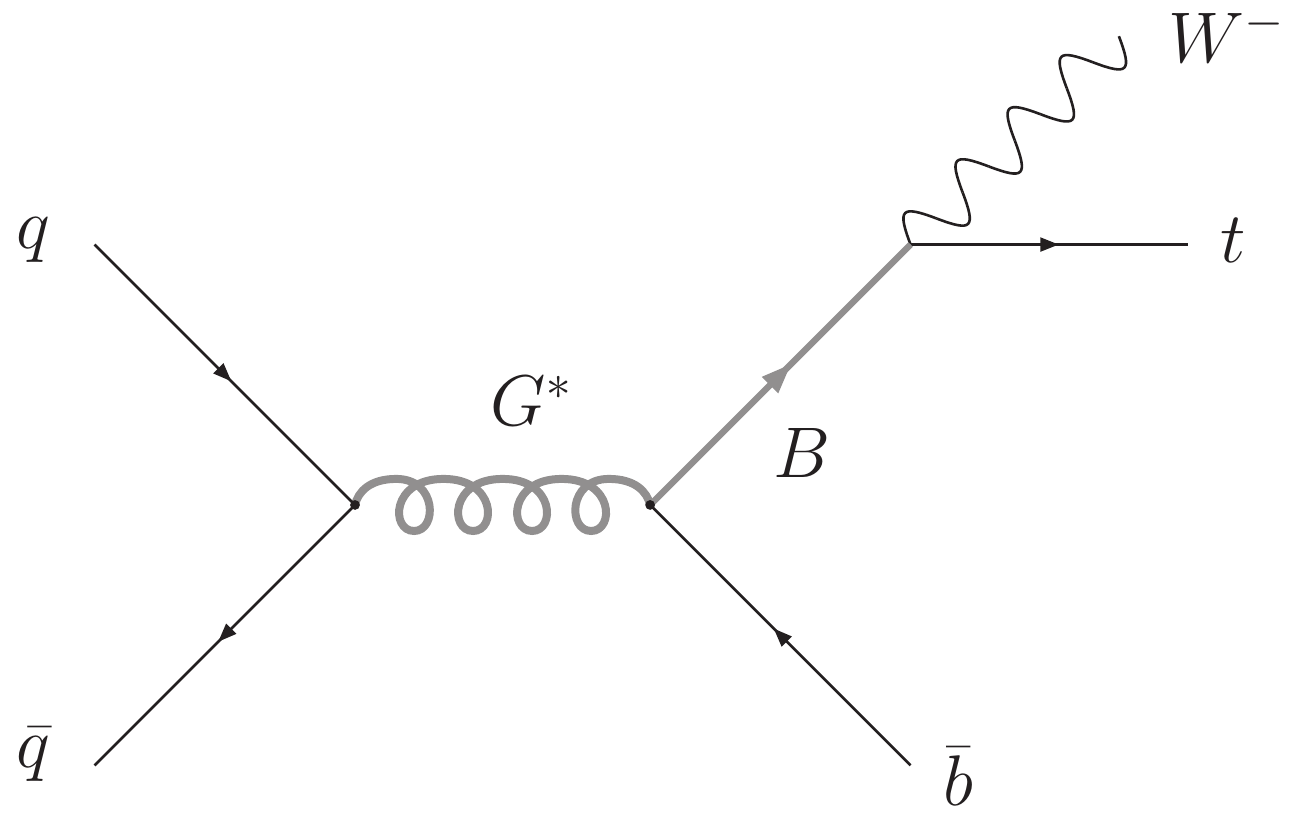}
\caption[]{
\label{fig:topologies}
\small
Signal topologies contributing to the $Wtb$ channel. The conjugate processes proceed through similar diagrams.
}
\end{center}
\end{figure}

There are  two signal topologies contributing to the $Wtb$ channel (see Fig.~\ref{fig:topologies}): either the $G^*$ decays to one SM and one 
heavy top ($\tilde T$), and the latter then decays to $Wb$; or the $G^*$ decays to one SM and one heavy bottom ($B$) and the 
latter decays to $Wt$.
%%
%%%%%%%%%%%%%%%%%%%%%%%%%%%%%%%%%%%%%%%%%%%%%%%%%%%%%%%%%%%%%%%%%%%%%%%
%%%%%%%%%%%%%%%%%%%%%%%%%%%%%%%%%%%%%%%%%%%%%%%%%%%%%%%%%%%%%%%%%%%%%%%
%%
Final states with one lepton thus follow from the decay chains:
\begin{equation} \label{eq:GtoTt}
\begin{split}
q\bar q & \to G^* \to \tilde T \bar t_R + c.c. \to W^+ b \bar t + c.c. \to l^\pm \nu \, b\bar b\, q\bar q \\[0.2cm]
q\bar q & \to G^* \to B \bar b_L  + c.c. \to W^- t \bar b + c.c. \to l^\pm \nu \, b\bar b\, q\bar q\, .
\end{split}
\end{equation}
The physical, observed process is $pp \to l^\pm \! + jets \, +  \not\!\! E_T$,
with up to four  jets from the hard scattering, two of which are $b$-jets. The same final state can however follow from another signal
topology, where $G^*$ decays to $t\bar t$:
\begin{equation}
q\bar q  \to G^* \to t \bar t \to W^+ b W^- \bar b \to l^\pm \nu \, b\bar b\, q\bar q\, .
\end{equation}
This is the decay chain on which previous studies aimed at the discovery of the $G^*$ focussed.

As a reference model for the signal we will consider the two-site model TS5 described in the previous section.
As explained there, since  the decay rates of $T \to W b_R$ and $\tilde B \to W t_L$ are strongly suppressed in the TS5 (see eq.(\ref{eq:fermionBRs})),
the decay chains $q\bar q\to G^* \to T \bar t + \tilde B \bar b +c.c. \to Wtb$ give a negligible contribution and will not be included in the following.
We perform our analysis adopting the following benchmark values of parameters ($m_h$ is the Higgs mass)
\begin{equation} \label{eq:benchmark}
\begin{aligned}
& Y_* = g_{*3} = 3\, , \quad \sin\varphi_{t_R} = 0.6 & & \text{(which implies: } \  \tan\theta_3 = 0.44 \, , \ \sin\varphi_{L} = 0.55\text{)}\\[0.3cm]
& \frac{M_{G^*}}{m_{\tilde T}} = 1.5 \, , & &   M_{G^*} = 1.5, \, 2.0,\,  3.0,\,  4.0 \, \text{TeV}\, , \quad m_h = 120\, \text{GeV}\, ,
\end{aligned}
\end{equation}
and we make a simple extrapolation to different values 
at the end. In particular, we fix the ratio of the $G^*$ mass over the heavy fermion mass to $M_{G^*}/m_{\tilde T} =1.5$, 
within the interval $m_{\tilde T} < M_{G^*} < 2 m_{T_{5/3}}$ where the processes of eq.(\ref{eq:GtoTt}) are kinematically allowed and fast.
For the reader's convenience,  we report  in Table~\ref{tab:benchmarkxsec}  the values of the $G^*$ production cross section 
at the  benchmark points of eq.(\ref{eq:benchmark}) (see also Fig.~\ref{fig:Gstarxsec}).
%
%%%%%%%%%%%%%%%%%%%%%%%%%%%%%%%%%%%%%%%%%%%%%%%%
\begin{table}
\begin{center}
\begin{tabular}{c|>{\hspace{15pt}}r@{}l>{\hspace{18pt}}r@{}l}
  \multirow{2}{*}{$M_{G^*} [\text{TeV}]$} & \multicolumn{4}{c}{$\sigma(pp \to G^*)$} \\[0.15cm]
  & \multicolumn{2}{c}{$\sqrt{s}=7\,$TeV} & \multicolumn{2}{c}{$\sqrt{s}=14\,$TeV}  \\[0.05cm]
\hline
  & &  & & \\[-0.3cm]
   $1.5$ &  $1.90\,$ &pb & $17.4\,$ &pb \\[0.1cm]
   $2.0$ &  $234\,$  &fb   & $4.23\,$ &pb \\[0.1cm]
   $3.0$ &  $4.12\,$ &fb  & $397\,$  &fb   \\[0.1cm]
   $4.0$ &  $0.05\,$ &fb  & $48.3\,$ &fb
\end{tabular}
\caption{\small 
Values of the $G^*$ production cross section at the LHC for the benchmark choice of parameters of eq.(\ref{eq:benchmark}).
\label{tab:benchmarkxsec}
}
\end{center}
\end{table}
%%%%%%%%%%%%%%%%%%%%%%%%%%%%%%%%%%%%%%%%%%%%%%%%
%
The particles' total decay widths are
\begin{equation}
\frac{\Gamma(G^*)}{M_{G^*}} = 0.052 \, , \qquad \frac{\Gamma(\tilde T)}{m_{\tilde T}} = 0.035\, , \qquad \frac{\Gamma(B)}{m_{B}} = 0.022\, , 
\end{equation}
and the branching ratios of the $Wtb$ and $t\bar t$ channels are~\footnote{Here and in the following
we use the short notation $G^* \to B b + \tilde T t$ to denote $G^* \to B\bar b + \tilde T \bar t +c.c.$.}
\begin{equation} \label{eq:benchmarkBR}
\begin{split}
BR(G^* \to  Bb \to Wtb) &= 0.17 \\[0.2cm]
BR(G^* \to \tilde T t \to Wtb) &= 0.08 
\end{split} \qquad
BR(G^* \to t\bar t) = 0.06\, .
\end{equation}
Hence, approximately two thirds (one third) of the $Wtb$ events that follow from a heavy-light decay of $G^*$
is from  $G^* \to B b$ ($G^* \to {\tilde T} t$).

The benchmark values of eq.(\ref{eq:benchmark}) have been chosen so as to have a large rate in the heavy-light channels.
This particular choice however also implies  a large branching ratio to a pair of light SM quarks, $BR(G^* \to q\bar q) = 0.22$ 
(see Fig.~\ref{fig:hldecays}), so that the dijet searches will also have a large discovery (or exclusion) power.~\footnote{In fact, 
the benchmark point (\ref{eq:benchmark}) might be already  excluded, although just marginally,  by the bounds from the dijet 
searches  reported by the CMS and ATLAS  collaborations~\cite{Chatrchyan:2011ns,Aad:2011fq} during the 
completion of this work. This is what suggests a naive rescaling of the bounds on  axigluons and colorons 
shown in Refs.~\cite{Chatrchyan:2011ns,Aad:2011fq}.  A more accurate derivation is however required to precisely determine the impact 
of the dijet searches on the parameter space of our  model. We leave this to a future study.}
As previously explained in section~\ref{subsec:Gstarpheno},  this
is not generic to the model, and it does not occur for other choices of the
parameters, like that of the left plot of Fig.~\ref{fig:hldecays2}, which 
implies large  branching ratios for the heavy-light channels and a very small one for $q\bar q$.

\subsection{Montecarlo simulation of signal and background}

We simulate the signal  by using MadGraph v4~\cite{MG-ME},
while for the background (see discussion below) we make use of both MadGraph and ALPGEN~\cite{Mangano:2002ea}.~\footnote{The 
choice of MadGraph to simulate the signal  is dictated by the possibility of easily implementing our model.
For the background, we have used ALPGEN to simulate  processes with a large number of particles from the hard
scattering. The factorization and renormalization scales have been set to be equal and chosen as follows:
$Q= M_{G^*}$ for the signal; $Q = \sqrt{m_W^2 + \sum_j p_{T j}^2}$ for $WWbb$; 
$Q = \sqrt{m_W^2 +  p_{T W}^2}$ for $Wbbj$,  $Wbbjj$, $Wbb3j$, $W3j$ and $W4j$.
}
In our parton-level analysis jets are identified 
with the quarks and gluons from the hard scattering.
If two quarks or gluons are closer than the separation $\Delta R =0.4$, they are merged into a single jet whose four-momentum is the
vectorial sum of the original momenta.
We require that the jets and the leptons satisfy  the following set of acceptance and isolation cuts:
\begin{equation}
\begin{aligned}
\ptj &\geq 30\, \gev  & \quad  | \etaj | &\leq 5    & \quad 
 \Delta R_{jj} &\geq 0.4 \\[0.2cm]
\ptl &\geq 20\, \gev & \quad  | \etal | &\leq 2.5  & \quad 
 \Delta R_{jl} &\geq 0.4 \, .
\end{aligned}
\label{eq:acceptance}
\end{equation}
Here $\ptj$ ($\ptl$) and $\etaj$ ($\etal$) are respectively 
the jet (lepton) transverse momentum and pseudorapidity, and 
$\Delta R_{jj}$, $\Delta R_{jl}$ denote the jet-jet and jet-lepton separations.
Jets that fail the $p_T$ cut of eq.(\ref{eq:acceptance}), as well as
leptons that fail the isolation and $p_T$ cut,  are  discarded and not considered in the reconstruction of the event.
$b$-jets are assumed to be correctly tagged with an efficiency $\eps_b = 0.6$ if their 
pseudorapidity satisfies $|\eta_b| \leq 2.5$.  We set the corresponding rejection rate on light jets to $\zeta_b = 100$.~\footnote{The rejection 
rate corresponds to the inverse of the probability of mistagging a light jet as a $b$-jet.}

Detector effects are roughly accounted for by performing a simple Gaussian smearing on the 
jet energy and momentum absolute value with $\Delta E/E = 100\%/\sqrt{E/\text{GeV}}$, and on the jet momentum direction using
an angle resolution $\Delta\phi =0.05$ radians and $\Delta\eta = 0.04$. Moreover, the missing energy $\etmiss$ of each event has been computed
by including a Gaussian resolution $\sigma(\etmiss) = a \cdot \sqrt{\sum_i E^i_T/\gev}$, where $\sum_i E^i_T$ is the scalar sum of the transverse
energies of all the reconstructed objects (electrons, muons and jets). We choose $a=0.49$.~\footnote{The numerical values of the above parameters,
as well as the $b$-tagging efficiency and rejection rate, have been chosen according to the performance of the ATLAS detector~\cite{Aad:2008zzm}.}

\begin{figure}[tbp]
\begin{center}
\includegraphics[width=0.485\textwidth,clip,angle=0]{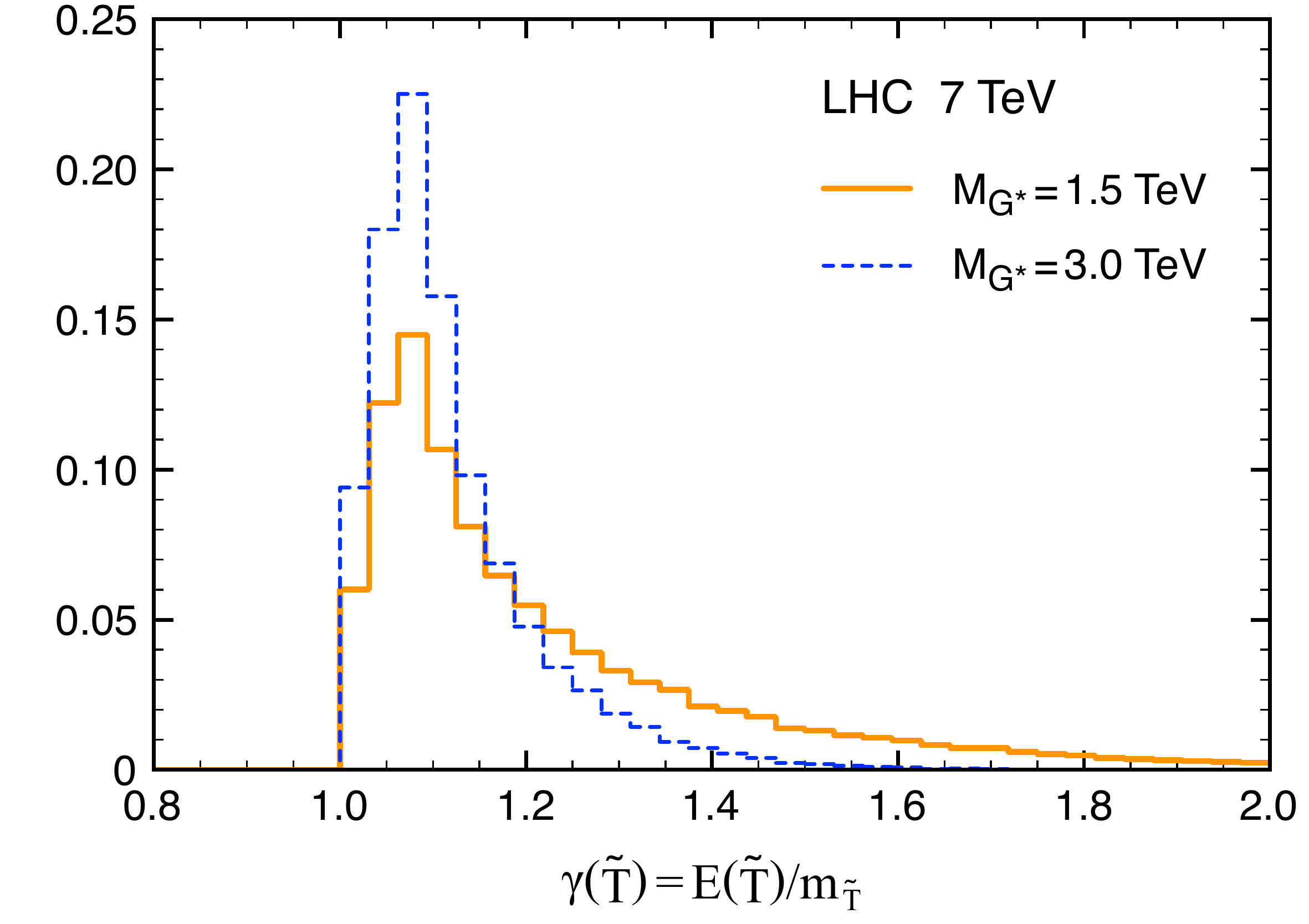}
\includegraphics[width=0.485\textwidth,clip,angle=0]{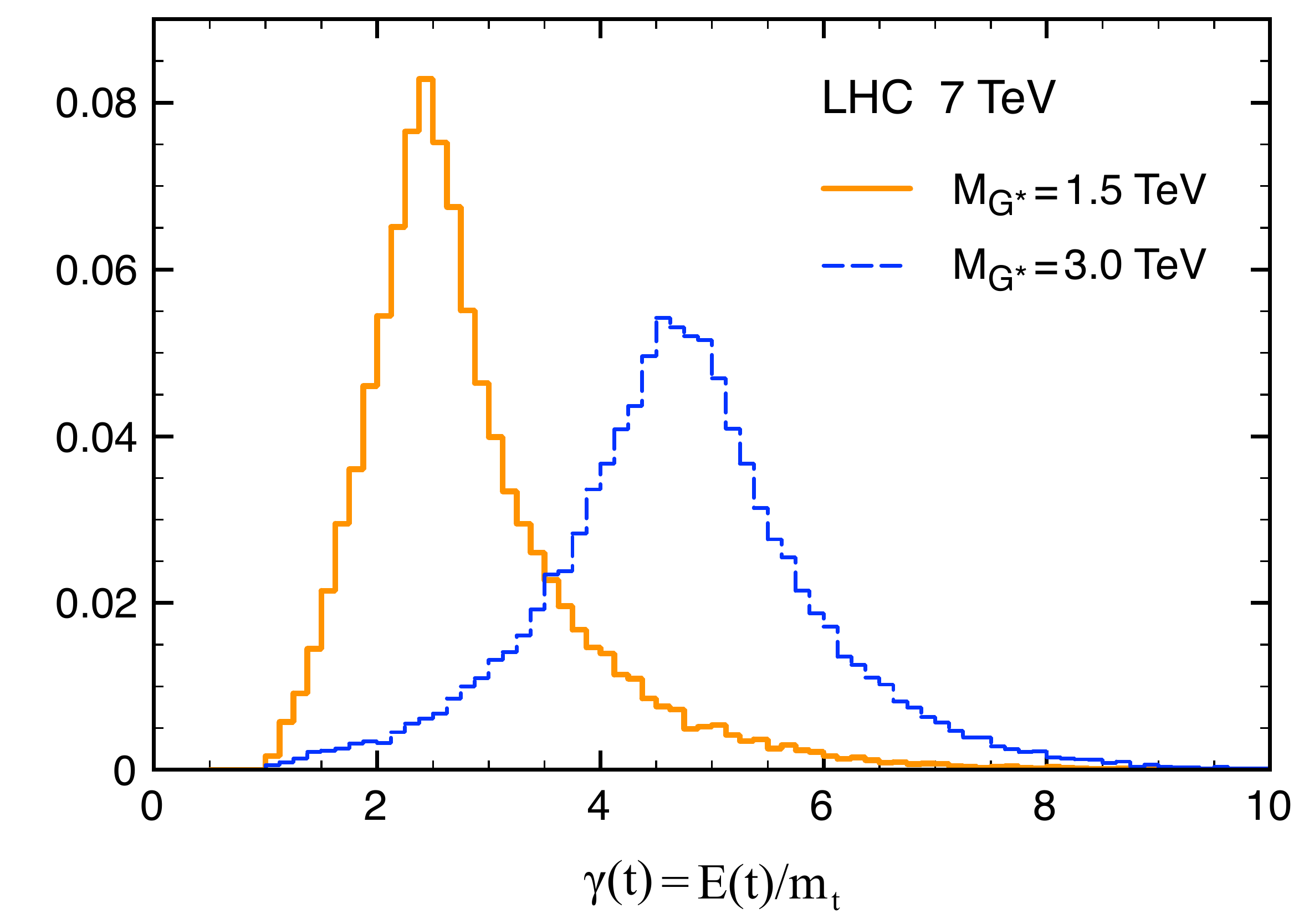}
\caption[]{
\label{fig:gamma}
\small
Distribution of the boost factor 
$\gamma =E/m$ of $\tilde T$ (left plot) and of the top quark (right plot) in signal events $G^* \to \tilde T t$ with 
$M_{G^*} = 1.5\,$TeV (continuous orange line) and $M_{G^*} = 3.0\,$TeV (dashed blue line) for $\sqrt{s} = 7\,$TeV.
Notice that having set $M_{G^*}/m_{\tilde T} = 1.5$ implies that $\tilde T$ is less boosted for $M_{G^*} = 3.0\,$TeV than 
for $M_{G^*} = 1.5\,$TeV. All the curves have been normalized to unit~area. 
}
\end{center}
\end{figure}

Table~\ref{tab:fractions} shows the fraction of   $G^* \to B b + \tilde T t$ events
where the jet content is reconstructed to be respectively $2j+2b$, $1j+ 2b$, $1j+1b$
(where $j$ denotes a light jet) for different values of the $G^*$ mass and $\sqrt{s}=14\,$TeV.
Very similar numbers hold for $\sqrt{s}=7\,$TeV.
%
%%%%%%%%%%%%%%%%%%%%%%%%%%%%%%%%%%%%%%%%%%%%%%%%
\begin{table}
\begin{center}
\begin{tabular}{c|ccc}
  $M_{G^*}$ & $2j + 2b$ & $1j + 2b$ & $1j + 1b$ \\[0.05cm]
\hline
  & & & \\[-0.3cm]
   $1.5\,$TeV & 0.42 & 0.31 (58\%) & 0.07 \\[0.1cm]
   $2.0\,$TeV & 0.29 & 0.42 (79\%) & 0.10 \\[0.1cm]
   $3.0\,$TeV & 0.13 & 0.52 (89\%) & 0.17 \\[0.1cm]
   $4.0\,$TeV & 0.07 & 0.53 (93\%) & 0.25 
\end{tabular}
\caption{\small 
Fraction of   $G^* \to Bb + \tilde T t$ events
where the jet content is reconstructed to be respectively $2j+2b$, $1j+ 2b$, $1j+1b$ 
(where $j$ denotes a light jet)  as a function of the $G^*$ mass for $\sqrt{s}=14\,$TeV.
In the third column,  the numbers in parenthesis show the relative fraction of $1j+ 2b$
events in which the light jet originates from the merging of the $q\bar q$ pair from the hadronic~$W$.  
\label{tab:fractions}
}
\end{center}
\end{table}
%%%%%%%%%%%%%%%%%%%%%%%%%%%%%%%%%%%%%%%%%%%%%%%%
%
In the third column,
the numbers in parenthesis show the relative fraction of $1j+ 2b$
events in which the light jet originates from the merging of the $q\bar q$ pair from the hadronic~$W$. 
As naively expected due to the increasing boost of all the decay products,  the number of events with four jets decreases for larger
$G^*$ masses, while the fraction of events in which the hadronic~$W$ is reconstructed as a single jet increases.
For this reason in our analysis {\em we will select events with at least three jets and exactly one lepton and two $b$-tags}:
\begin{equation} \label{eq:evsel}
pp \to l^\pm \! + n\, jets \, +  \not\!\! E_T\, , \qquad n\geq 3 \, , \ \ \ \text{2 $b$-tags}
\end{equation}
where all objects must satisfy the acceptance and isolation cuts of eq.(\ref{eq:acceptance}).
Most likely, the use of boosted jet techniques~\footnote{See for example the report prepared for the BOOST2010 workshop~\cite{boost}
for a review and a complete list of references.}
can lead to a better sensitivity on events with  three jets, and
allow the study of the extreme case, most probable at high $G^*$ masses, in which the hadronically decayed top is reconstructed as a single fat jet.
Notice, on the other hand, that while the SM  (top or bottom) quark originating from the $G^*$ decay is always highly boosted,
for $m_{G^*}/m_{\tilde T} =1.5$ the heavy quark is not. This is clearly shown by Fig.~\ref{fig:gamma} in the case of $G^* \to \tilde T t$ events.
%%
%%%%%%%%%%%%%%%%%%%%%%%%%%%%%%%%%%%%%%%%%%%%%%%%%%%%%%%%%%%%%%%%%%%%%%%
%%%%%%%%%%%%%%%%%%%%%%%%%%%%%%%%%%%%%%%%%%%%%%%%%%%%%%%%%%%%%%%%%%%%%%%
%%
As a consequence, an event selection strategy which relies on a large boost of \textit{all} the decay products, adopted for example by some 
of the LHC searches for heavy resonances decaying to $t\bar t$, might have a poor efficiency on our signal.

The largest SM background after the event selection of eq.(\ref{eq:evsel}) is the irreducible background $WWbb$, 
which includes the resonant sub-processes $Wtb\to WWbb$ (single top) and $t\bar t \to WWbb$.
The latter, in particular, gives the largest contribution.
We have simulated the $WWbb$ events by using MadGraph.
Another background which will turn out to be important after imposing our full set of kinematic cuts is $Wbb+jets$. 
We have simulated $Wbbj$,  $Wbbjj$ and $Wbb3j$ (this latter process only for $\sqrt{s}=14\,$TeV)  using ALPGEN.
Including all these samples with increasing multiplicity of light jets in the final state is redundant,
and in principle leads to a double counting of kinematic configurations.
A correct procedure would be resumming soft and collinear emissions by means of a parton shower,
and adopting some matching technique to avoid double counting.
For simplicity, in our analysis we retain all the $Wbb+ n\, jets$ samples; in this way we expect to obtain a conservative estimate
of the background. Notice also that some of the cuts we will impose tend to suppress
the events with larger number of jets and thus to reduce the amount of double counting.
Finally, the last background that we have considered is $W+jets$, where two light jets are mistagged as $b$-jets.
We have generated the processes $W3j$ and $W4j$ with ALPGEN.
As for the $Wbb+ jets$ background, we conservatively include both these sample.
In this case, however, the issue of double counting can  be safely ignored since 
the $W+jets$ background will turn out to be much smaller than the others at the end of our analysis.

We did not include other reducible backgrounds which are expected to be subdominant, in particular: $b\bar b+jets$
where one light jets is misreconstructed as a lepton (it should be possible to reduce it
down to a negligible level by requiring enough missing energy in the event); single-top processes $t+jets$,
$tb+jets$, $Wt+jets$ (after the request of two $b$-tags all these are expected to be much smaller than the 
single-top background $Wtb$,~\footnote{See for example Table~32 at page 32 of Ref.~\cite{Mangano:2002ea}.}
which is included in our analysis).

While our analysis is carried out at the parton level, we expect it to be robust against the inclusion of detector and 
showering effects. 
Our simple Gaussian smearing of the jets' energy and momentum should correctly reproduce the main impact of detector effects
on our event selection and  reconstruction. 
The request of two $b$-tags and the kinematic cuts that we will impose (like for example on the invariant mass of the hadronic $W$, 
see next section) should  reduce the effect of extra radiation in the signal.
A detailed study of initial and final state radiation, underlying event and multiple parton collisions is beyond the scope of the present paper
and is left for future analyses.

\subsection{Event selection}

The second column of Tables~\ref{tab:cutflow7TeV1}  and~\ref{tab:cutflow14TeV1}  reports the value of the cross section for the signal and the 
main SM backgrounds after the selection (\ref{eq:evsel}) based on the acceptance and isolation cuts of eq.(\ref{eq:acceptance}), 
respectively for $\sqrt{s} = 7\,$TeV and $14\,$TeV.
%
%%%%%%%%%%%%%%%%%%%%%%%%%%%%%%%%%%%%%%%%%%%%%%%%%
\begin{table}
\begin{center}
{\small
\begin{tabular}{|ll|c|c|c|}
\hline
 & & & & \\[-0.4cm]
 \multicolumn{2}{|c|}{\textsf{LHC $\mathsf{\sqrt{s} = 7\,}$TeV}} & acceptance & $\nu\,+\,$top rec. & zero-cost  \\[0.1cm]
\hline
 & & & & \\[-0.3cm]
\multirow{2}{*}{$M_{G^*}=1.5\,$TeV} 
 & $G^*\to\tilde Tt + Bb$  & 29.8   & 23.8 & 22.2  \\[0.1cm]
 & $G^*\to t \bar t$            & 5.85   & 4.72 & 3.21  \\[0.25cm]
\multirow{2}{*}{$M_{G^*}=2.0\,$TeV} 
 & $G^*\to\tilde Tt + Bb$  & 3.29    & 2.59 & 2.51  \\[0.1cm]
 & $G^*\to t \bar t$            & 0.71   & 0.56 & 0.34  \\[0.25cm]
\multirow{2}{*}{$M_{G^*}=3.0\,$TeV} 
 & $G^*\to\tilde Tt + Bb$  & 0.04    & 0.03 & 0.03  \\[0.1cm]
 & $G^*\to t \bar t$            & 0.06   & 0.05 & 0.02  \\[0.35cm]
 & $WWbb$            & 4838 & 3932 & 167  \\[0.25cm]
 & $Wbbj$              & 210   & 156  & 7.78  \\[0.25cm]
 & $Wbbjj$             & 102   & 67.5 & 2.17  \\[0.25cm]
 & $W3j$                & 18.8  & 14.5 & 0.83  \\[0.25cm]
 & $W4j$                & 8.89  & 6.36 & 0.31  \\[0.35cm]
 & Total                  & & & \\
 & background       & 5177 & 4177 & 179   \\[0.15cm]
\hline
\end{tabular}
}
\caption{
\label{tab:cutflow7TeV1}
\small 
Cross sections, in fb, at $\sqrt{s}=7\,$TeV for the signal and the main backgrounds after 
the selection~(\ref{eq:evsel}) based on the acceptance cuts of eq.(\ref{eq:acceptance}) (second column); 
after the neutrino and top quark reconstruction (third column); after imposing the `zero-cost' cuts of eq.(\ref{eq:zerocost7TeV}) 
(fourth column). For each channel, the proper branching ratio to a one-lepton final state has been included.
}
\end{center}
\end{table}
%%%%%%%%%%%%%%%%%%%%%%%%%%%%%%%%%%%%%%%%%%%%%%%%%
\begin{table}
\begin{center}
{\small
\begin{tabular}{|ll|c|c|c|}
\hline
 & & & & \\[-0.4cm]
 \multicolumn{2}{|c|}{\textsf{LHC $\mathsf{\sqrt{s} = 14\,}$TeV}} & acceptance &  $\nu\,+\,$top rec. & zero-cost  \\[0.1cm]
\hline
 & & & & \\[-0.3cm]
\multirow{2}{*}{$M_{G^*}=1.5\,$TeV} 
 & $G^*\to \tilde Tt + Bb$  & 262  & 209  & 192 \\[0.1cm]
 & $G^*\to t \bar t$            & 42.6 & 34.8 & 23.7 \\[0.25cm]
\multirow{2}{*}{$M_{G^*}=2.0\,$TeV} 
 & $G^*\to \tilde Tt + Bb$  & 56.9 & 45.0 & 42.7 \\[0.1cm]
 & $G^*\to t \bar t$            & 7.00 & 5.76 & 3.93 \\[0.25cm]
\multirow{2}{*}{$M_{G^*}=3.0\,$TeV} 
 & $G^*\to\tilde Tt + Bb$  & 3.80  & 2.95 & 2.91 \\[0.1cm]
 & $G^*\to t \bar t$            & 0.46 & 0.38 & 0.22 \\[0.25cm]
\multirow{2}{*}{$M_{G^*}=4.0\,$TeV} 
 & $G^*\to \tilde Tt + Bb$  & 0.32 & 0.24 & 0.24 \\[0.1cm]
 & $G^*\to t \bar t$            & 0.08 & 0.07 & 0.03 \\[0.3cm]
 & $WWbb$            & 27671 & 22383 & 724 \\[0.25cm]
 & $Wbbj$              & 794    &  573    & 25.9 \\[0.25cm]
 & $Wbbjj$             & 574    &  354    & 9.30 \\[0.25cm]
 & $Wbb3j$            & 215    &  119    & 0.63 \\[0.25cm]
 & $W3j$                & 67.6   &  51.3   & 2.90 \\[0.25cm]
 & $W4j$                & 41.2   &  28.2   & 1.15 \\[0.25cm]
 & Total                  & & &  \\
 & background       & 29363 & 23509 & 764  \\[0.15cm]
\hline
\end{tabular}
}
\caption{
\label{tab:cutflow14TeV1}
\small 
Cross sections, in fb, at $\sqrt{s}=14\,$TeV for the signal and the main backgrounds after 
the selection~(\ref{eq:evsel}) based on the acceptance cuts of eq.(\ref{eq:acceptance}) (second column); 
after the neutrino and top quark reconstruction (third column); after imposing the `zero-cost' cuts of eq.(\ref{eq:zerocost14TeV}) 
(fourth column). For each channel, the proper branching ratio to a one-lepton final state has been included.
}
\end{center}
\end{table}
%%%%%%%%%%%%%%%%%%%%%%%%%%%%%%%%%%%%%%%%%%%%%%%%%
% %
One can see that at this stage the background dominates by far over the signal.
We can however exploit the peculiar kinematics of the signal to perform a first set of cuts and reduce the background to 
a much smaller level without touching the signal.

\begin{figure}[tbp]
\begin{center}
\includegraphics[width=0.485\textwidth,clip,angle=0]{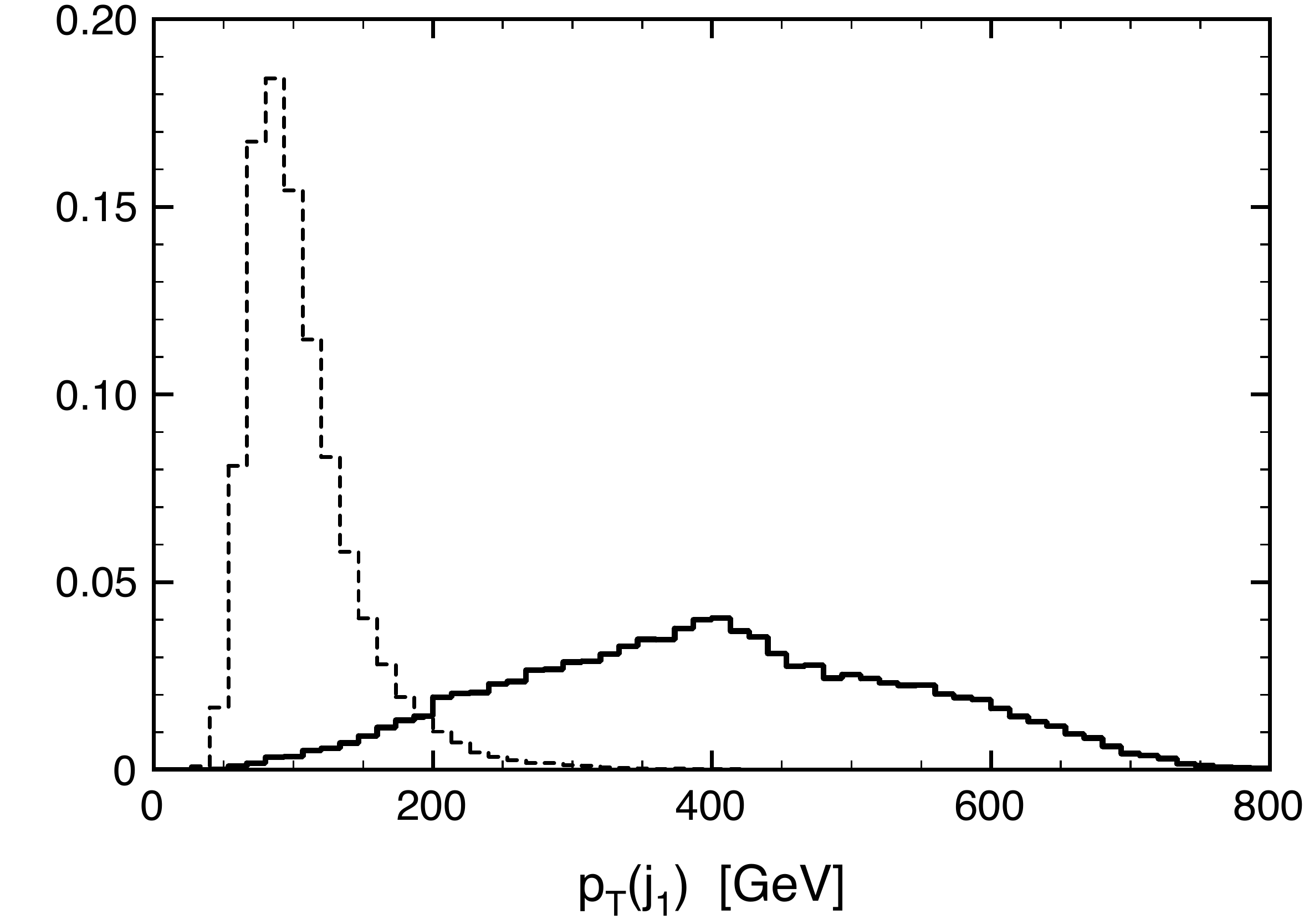}
\includegraphics[width=0.485\textwidth,clip,angle=0]{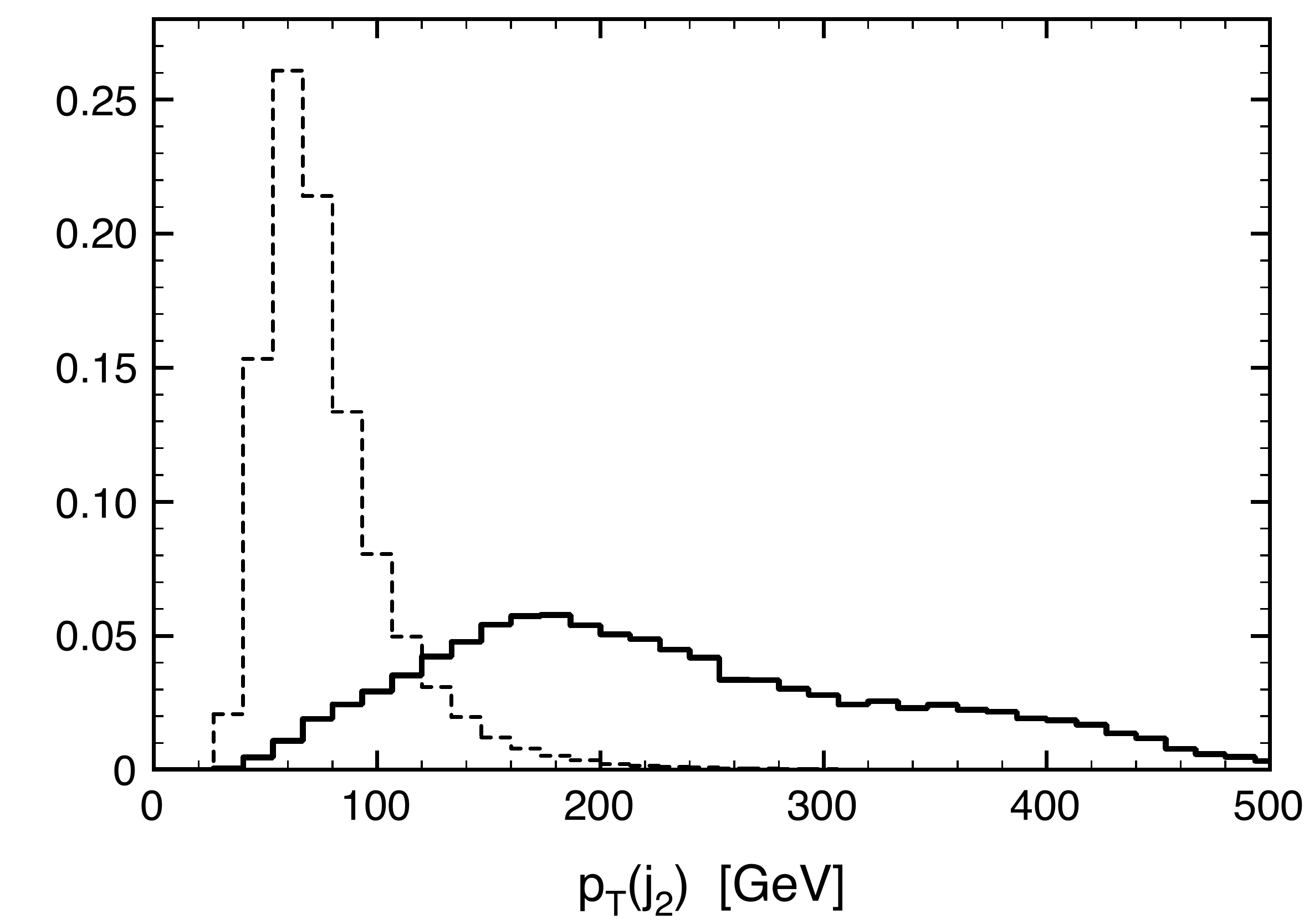}
\caption[]{
\label{fig:pTj12}
\small
Differential distribution of the $p_T$ of the leading jet (left plot) and next-to-leading jet (right plot)  
after the acceptance cuts of eq.(\ref{eq:acceptance}) for $\sqrt{s} =7\,$TeV.
The continuous line and the dashed line respectively show the signal at $m_{G^*}=1.5\,$TeV and the total background.
All the curves have been normalized to unit~area. 
}
\end{center}
\end{figure}

One basic feature of the signal is that the lepton and the jets in the event
tend to be very energetic,  as they are the final products of the decay chain of a new heavy particle (either the $G^*$ or the heavy fermion).
This is evident for example from the plots of Fig.~\ref{fig:pTj12}, which show the $p_T$ of the leading jet ($j_1$) and of
the next-to-leading jet ($j_2$) for the signal at $M_{G^*}=1.5\,$TeV and for the total background.
%
%%%%
%%%%
%
We will impose a cut on $p_T(j_{1,2})$ in the following, see eqs.(\ref{eq:zerocost7TeV}),(\ref{eq:zerocost14TeV}).

In order to fully exploit the specific topology of the signal, however,
it is convenient 
to reconstruct the intermediate $Wtb$ state that follows from the decay of the $G^*$ and of the heavy fermion, see Fig.~\ref{fig:topologies}.
To do this, we first reconstruct the longitudinal momentum of the neutrino, up to a twofold ambiguity,
by enforcing the on-shell condition
$m(l\nu) = m_W$.  Events where the second-order equation has no (real) solution are 
removed.~\footnote{Although this algorithm is very rough and does not account for the amount of off-shellness of the $W$ boson,
it has a sufficiently large efficiency on our signal: $\eps_\nu \sim 0.8$. We do not attempt here to adopt more refined strategies,
which are however well known and can be easily implemented in a full analysis.}
We thus have two pairs $(l\nu)$, one for each neutrino solution, which represent our two candidates 
for the first $W$ in the signal:  $W_{l_{1,2}} = (l\nu_{1,2})$.
Next, we use the fact that in the signal there is a second $W$ that decays hadronically, and label all the  jets
other than the two $b$-jets  to form our hadronic $W$ candidate, $W_h$.
Notice that in events with exactly three hard jets, $W_h$ will consist of a single jet.
Starting from the three $W$ candidates and the two $b$-jets in the event, there are six  $(Wb)$ pairs that one can form. 
We select the pair whose invariant mass is closest to the top quark mass, and label it as our top quark candidate: $(W_t b_t)$.
The remaining $b$ and $W$ candidates will be labeled as $b_{\nottop}$ and $W_{\nottop_{1,2}}$, since in the signal they do not come 
from the decay of a top quark. In the case in which the $W$ selected as belonging to the top quark is one of the two leptonic candidates,
$W_t = W_{l_i}$, the other leptonic candidate is discarded. In other words, the top reconstruction in this case 
gives us a criterion to select one of the two neutrino solutions. If instead the $W$ selected as belonging to the top quark is
the hadronic one, $W_t = W_h$, then both the remaining $W$ candidates are kept.
As a first loose cut on the invariant mass of the reconstructed top we require $80\,\text{GeV} < m(W_t b_t) < 250\,$GeV. 
The efficiency of this cut on the  signal and on the backgrounds with at least one top quark is $\sim 100\%$.
The value of the cross sections after both the neutrino and  top reconstruction are reported in the third column of
Tables~\ref{tab:cutflow7TeV1}  and~\ref{tab:cutflow14TeV1}.

\begin{figure}[tbp]
\begin{center}
\includegraphics[width=0.485\textwidth,clip,angle=0]{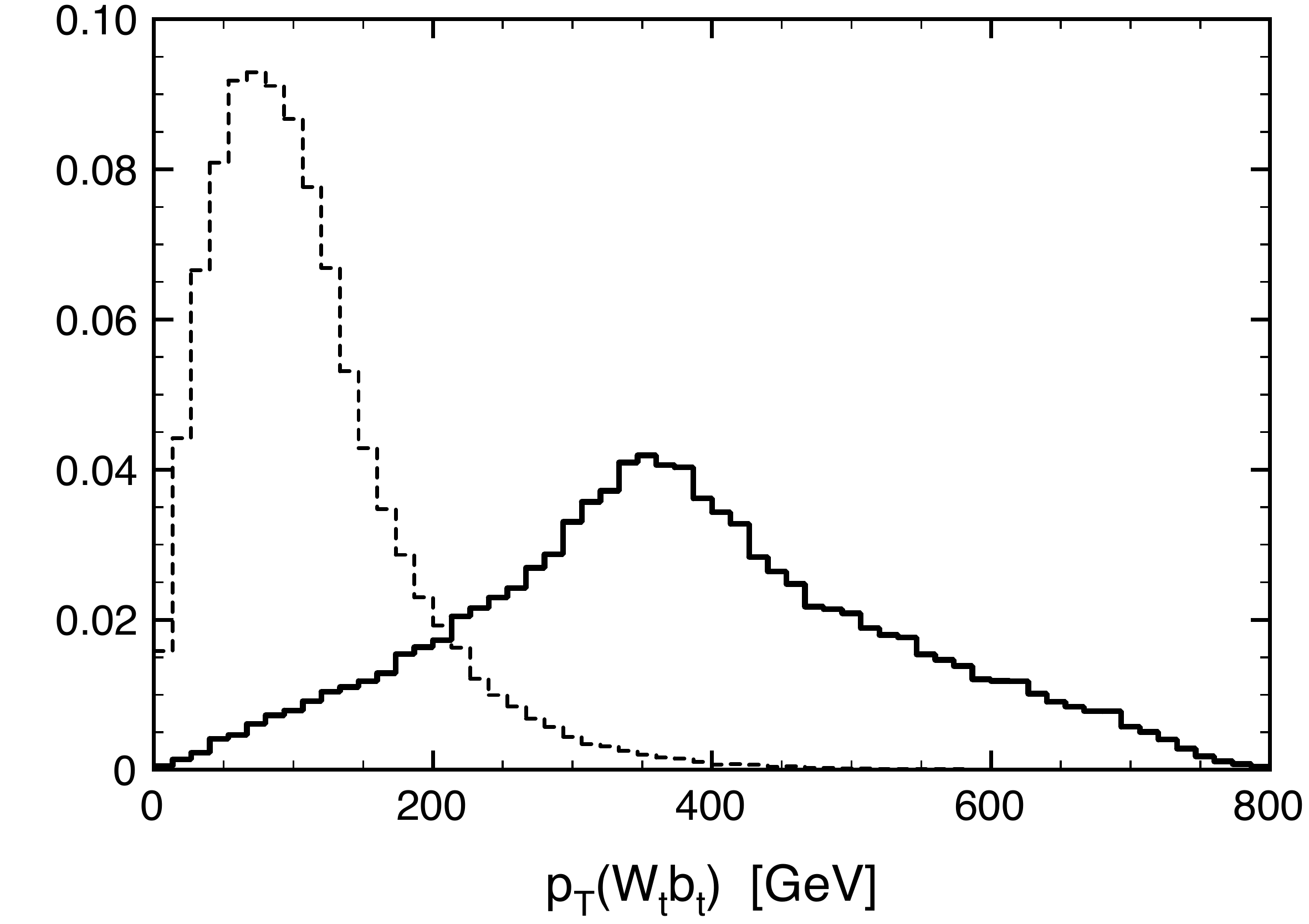}
\includegraphics[width=0.485\textwidth,clip,angle=0]{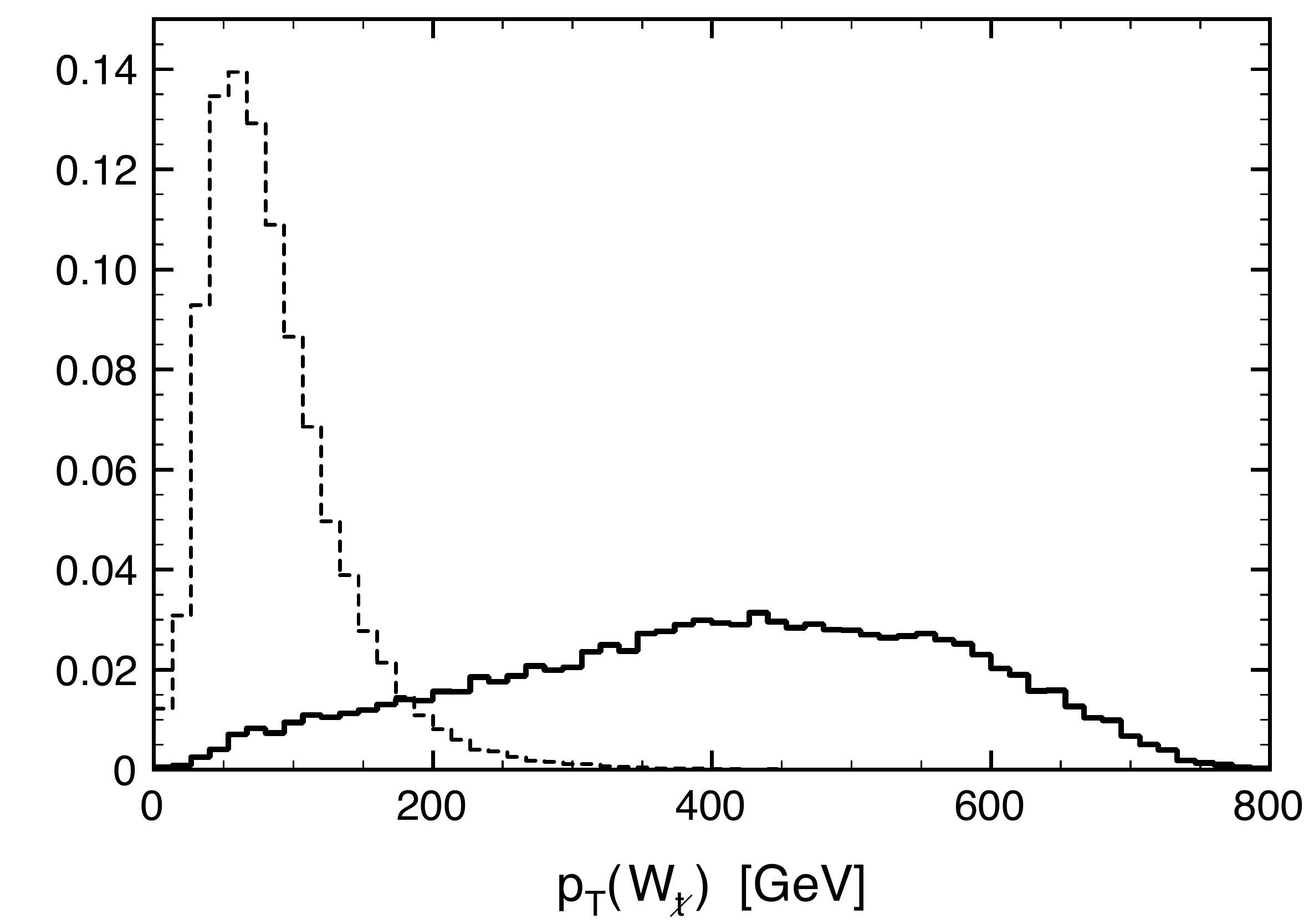}
\\[0.3cm]
\includegraphics[width=0.485\textwidth,clip,angle=0]{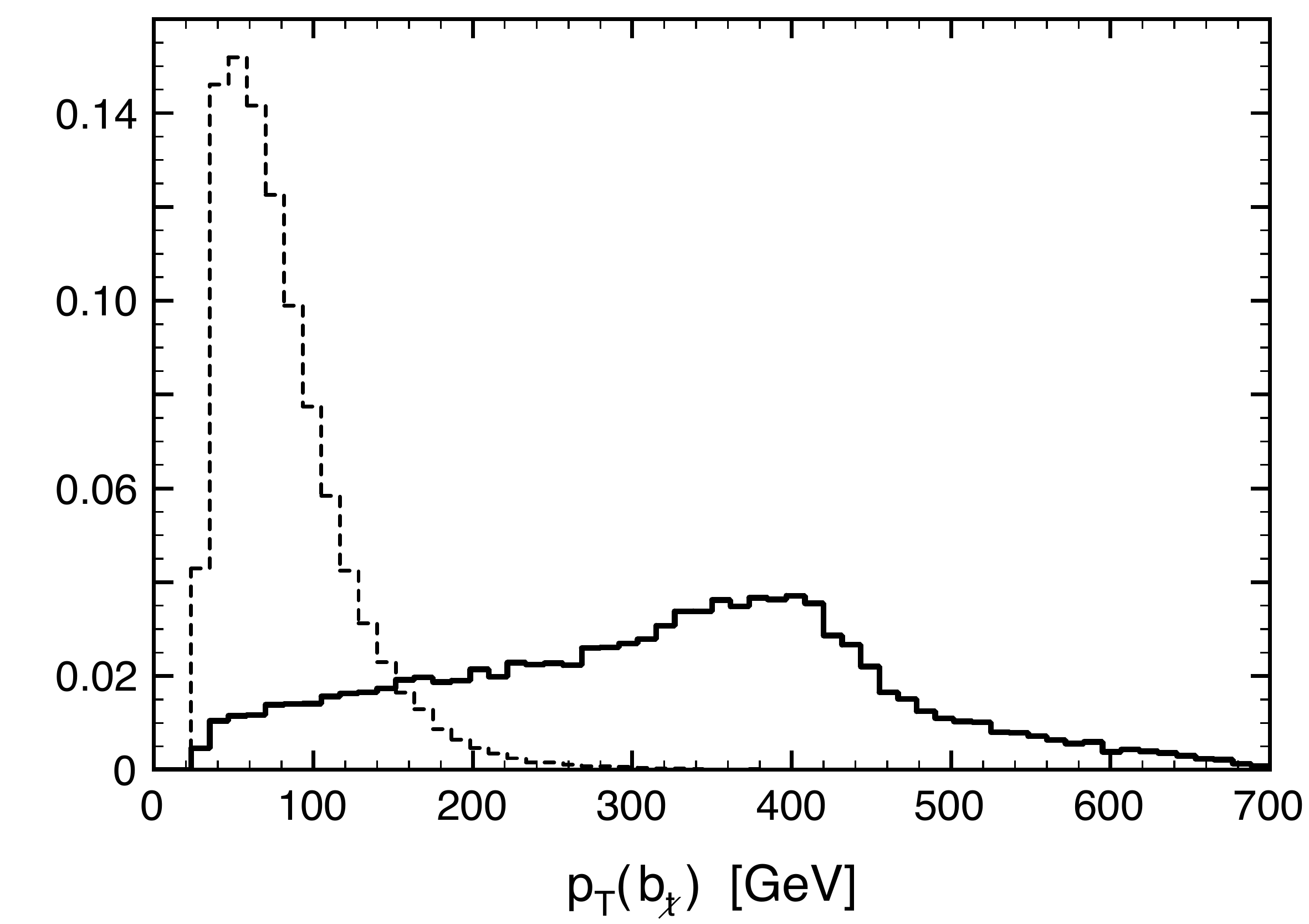}
\includegraphics[width=0.485\textwidth,clip,angle=0]{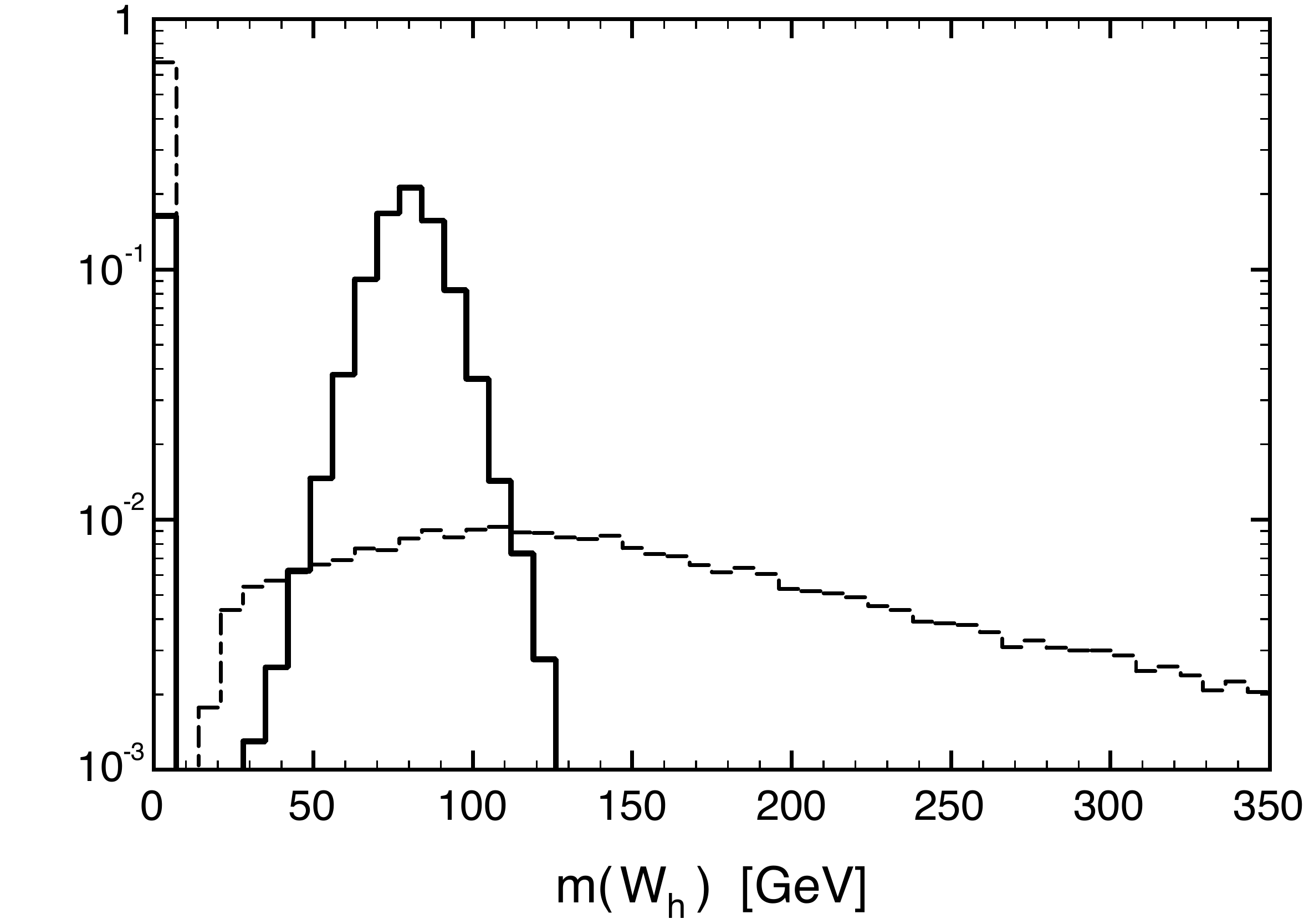}
\caption[]{
\label{fig:pTtpTWpTb}
\small
Differential distributions after the neutrino and  top reconstruction for $\sqrt{s} =7\,$TeV.
Upper left plot: $p_T$ of the top candidate, $p_T(W_t b_t)$; Upper right plot: $p_T$ of the $W$ candidate
labeled as not belonging to the top quark, $p_T(W_\nottop)$ (if two such candidates exist they have the same transverse 
momentum by construction, see text); Lower left plot: $p_T$ of the $b$-jet labeled as not belonging to the top quark, $p_T(b_\nottop)$;
Lower right plot: invariant mass of the hadronic $W$ candidate, $m(W_h)$.
In this latter plot, the first bin is populated by events with exactly three jets, for which $m(W_h)=0$ in our partonic analysis.
The continuous line shows the signal at $m_{G^*}=1.5\,$TeV; in the upper plots and the lower left plot, the dashed line
shows the total background; in the lower right plot, the dashed line
shows the (sum of the) backgrounds without a second $W$, namely $Wbb+jets$ and $W+jets$.
All the curves have been normalized to unit~area. 
}
\end{center}
\end{figure}

The identification of the $Wtb$ intermediate state allows us to fully take advantage of the peculiarity of the signal, 
where each of these three SM particles is extremely energetic.
This is illustrated by Fig.~\ref{fig:pTtpTWpTb}, which shows the distributions of the 
transverse momenta of the reconstructed top, $p_T(W_t b_t)$,
and of the $W$ candidate and $b$-jet not belonging to it, $p_T(W_\nottop)$, $p_T(b_\nottop)$, at $\sqrt{s} = 7\,$TeV.
Also shown is the distribution of the invariant mass of the hadronic $W$ candidate, $m(W_h)$.
Similar plots are obtained for $\sqrt{s} = 14\,$TeV.
%%
%%%%%%%%%
%%%%%%%%%
%%
As a first set of cuts, we thus use the kinematic observables of Figs.~\ref{fig:pTj12},~\ref{fig:pTtpTWpTb}.
At $\sqrt{s} = 7\,$TeV we require:
\begin{equation} \label{eq:zerocost7TeV}
\begin{aligned}
p_T(j_1) & \geq 155\,\gev \qquad  & p_T(j_2) & \geq 75\,\gev \qquad  & m(W_h) & \leq 200\,\gev \\[0.3cm]
p_T(W_t b_t) &\geq 105 \,\gev \qquad & p_T(W_{\nottop}) & \geq 90 \,\gev \qquad & p_T(b_\nottop) &\geq 65 \,\gev\, ,
\end{aligned}
\end{equation}
while at $\sqrt{s} = 14\,$TeV we impose slightly stronger cuts as follows:
\begin{equation} \label{eq:zerocost14TeV}
\begin{aligned}
p_T(j_1) & \geq 175\,\gev \qquad  & p_T(j_2) & \geq 85\,\gev \qquad  & m(W_h) & \leq 200\,\gev \\[0.3cm]
p_T(W_t b_t) &\geq 110 \,\gev \qquad & p_T(W_{\nottop}) & \geq 110 \,\gev \qquad & p_T(b_\nottop) &\geq 70 \,\gev\, .
\end{aligned}
\end{equation}
In each case, the numerical values of the cuts have been chosen so that each cut individually reduces the signal $G^* \to \tilde Tt + Bb$
at $M_{G^*}=1.5\,$TeV by no more than $\sim 2\%$, so that they are basically at `zero cost' for the signal.
Notice that when two $W_{\nottop}$ candidates exist, by construction they  have the same transverse momentum,
so that the cut on $p_T(W_{\nottop})$ of eqs.(\ref{eq:zerocost7TeV}),(\ref{eq:zerocost14TeV}) applies to either of them.
As clearly illustrated by Fig.~\ref{fig:pTtpTWpTb}, the cut on $m(W_h)$ is useful to reduce the backgrounds that do not have a second $W$, 
in particular $Wbb+jets$ and $W+jets$.

\begin{figure}[tbp]
\begin{center}
\includegraphics[width=0.6\textwidth,clip,angle=0]{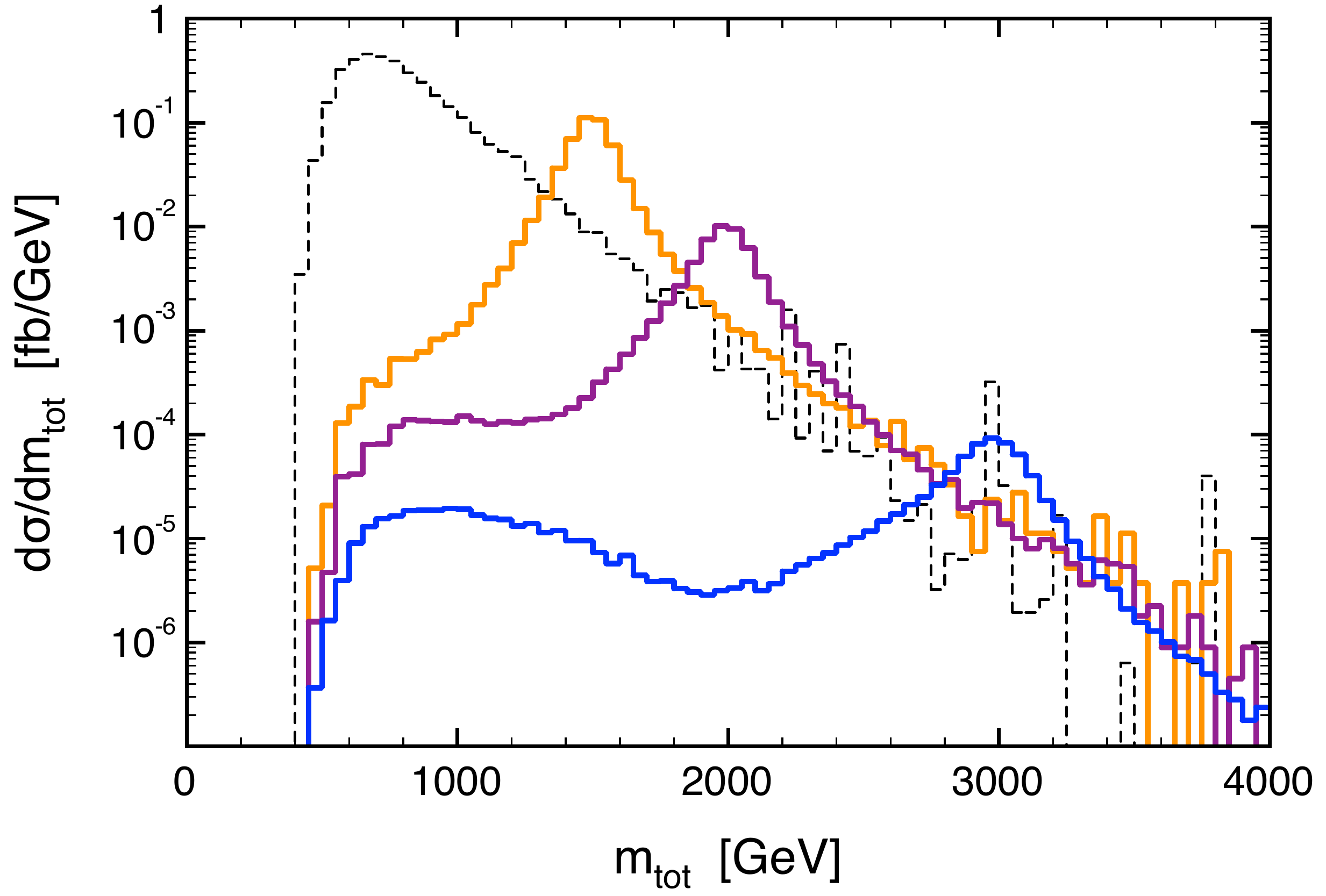}
\caption[]{
\label{fig:mTOT}
\small
Differential cross section as a function of the total invariant mass of the $Wtb$ system, 
$m_{tot} \equiv m(W_t b_t W_\nottop b_\nottop)$, after the  cuts at `zero cost' of eq.(\ref{eq:zerocost7TeV}) at $\sqrt{s}=7\,$TeV.
The dashed curve denotes the total background; the orange, purple, and blue 
continuous curves denote the signal respectively at $M_{G^*} = 1.5, 2.0, 3.0\,$TeV.
The long tail of the signal curves at low $m_{tot}$ is due to events with an off-shell $G^*$.
When two $W_l$ candidates exist in an event, the corresponding two solutions for $m_{tot}$ have been both included, each with weight $1/2$.
}
\end{center}
\end{figure}

\begin{figure}[tb!]
\begin{center}
\includegraphics[width=0.6\textwidth,clip,angle=0]{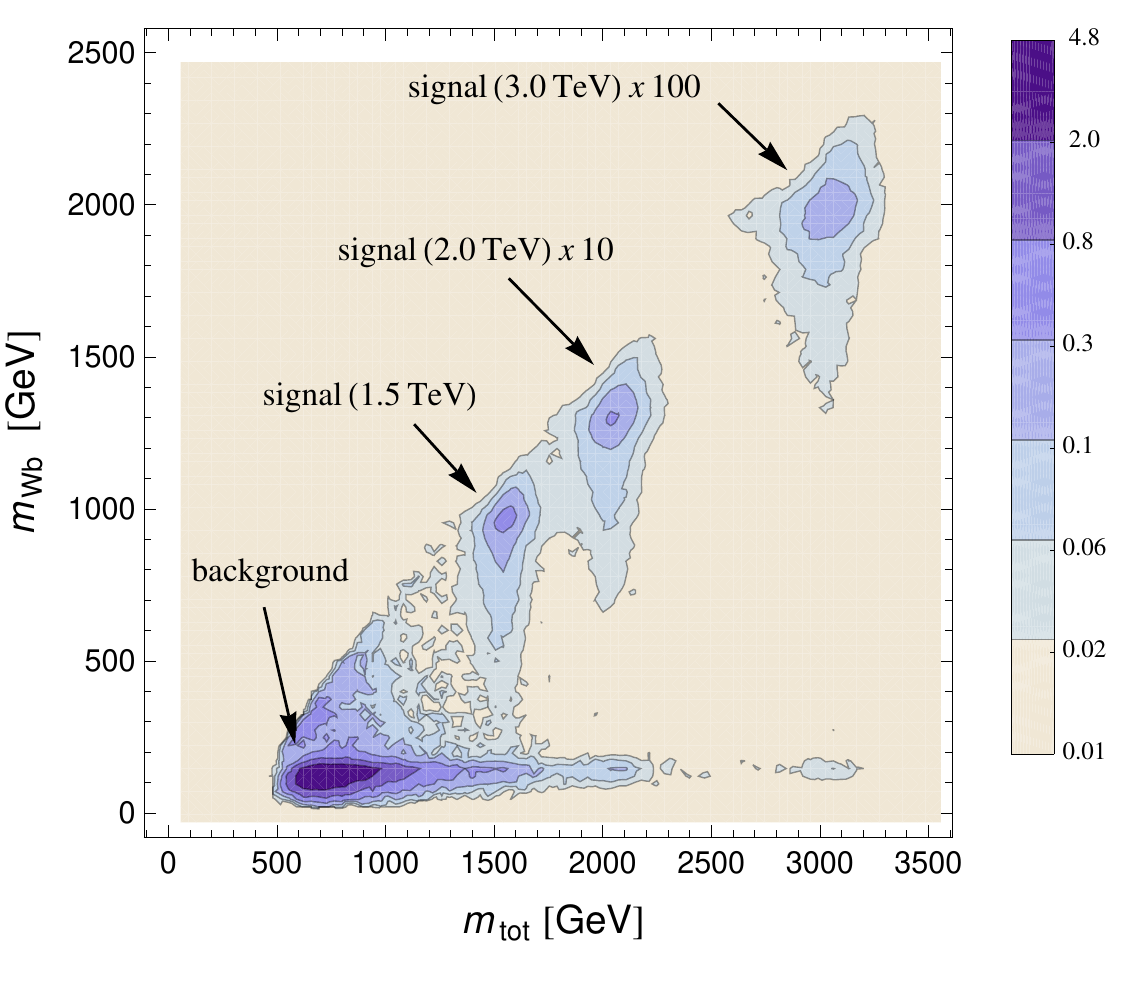}
\caption[]{
\label{fig:mtotmWb}
\small
Contour plot of the signal and total background cross sections after the cuts at `zero cost'  of eq.(\ref{eq:zerocost7TeV}) in the plane $(m_{tot}, m_{Wb})$, 
where $m_{Wb} \equiv m(W_\nottop b_\nottop)$ and $m_{tot} \equiv m(W_\nottop b_\nottop W_t b_t)$. The center-of-mass energy is $\sqrt{s}=7\,$TeV.
Different colors correspond to areas of different values of the doubly differential cross section $d^2\sigma/dm_{Wb}dm_{tot}$, 
as reported in the vertical color key on the right in units $\text{ab}/\gev^2$.
The signals at $M_{G^*} =2.0\,$TeV and $3.0\,$TeV have been rescaled respectively by a factor 10 and 100. 
}
\end{center}
\end{figure}

The cross sections of the signal and the backgrounds after the cuts at `zero cost' of eq.(\ref{eq:zerocost7TeV}) and (\ref{eq:zerocost14TeV})
are reported  in the fourth column of respectively Table~\ref{tab:cutflow7TeV1}  and~\ref{tab:cutflow14TeV1}.  At this stage, although the 
background has been strongly reduced, 
it is still dominant over the signal. In particular, the largest background comes from the resonant contribution  $t\bar t\to WWbb$. 
An obvious strategy to suppress it, adopted  in previous analyses aimed at uncovering the signal  $G^*\to t\bar t$,
is that of  cutting on the total invariant mass of the event. 
This is illustrated by Fig.~\ref{fig:mTOT}, which shows the differential cross sections
of the signal and of the total background for $\sqrt{s} = 7\,$TeV as functions of the invariant mass of the $Wtb$ system after the 
cuts of eq.(\ref{eq:zerocost7TeV}).
%%
%%%%%%%%%%%%%%%%%%%%%%%%%%%%%%%%%%%%%%%%%%%%%%%%%%%%%%%%%%%%%%%%%%%%%%%
%%%%%%%%%%%%%%%%%%%%%%%%%%%%%%%%%%%%%%%%%%%%%%%%%%%%%%%%%%%%%%%%%%%%%%%
%%
%
Obtaining a better significance of the $G^*\to t\bar t$ signal over the background, especially at 
large $G^*$ masses and widths,  requires  additional and more sophisticated tools, like for the example the use of 
a left-right polarization asymmetry~\cite{Agashe:2006hk,Lillie:2007yh}.
The peculiar topology of the process $G^* \to \tilde T t + Bb$, on the other hand, 
suggests a further simple strategy, namely requiring that the invariant mass of the $(W_\nottop b_\nottop)$ system be 
much bigger than the top mass $m_t$. 
In the case of $G^*\to \tilde Tt$, indeed,  $m(W_\nottop b_\nottop)$ peaks at  $m_{\tilde T}$,
and even in $G^*\to Bb$ events it tends to be much larger than $m_t$.
This is shown by the contour plot
of Fig.~\ref{fig:mtotmWb}, which reports the isocurves of the (doubly differential) cross section in the plane
$(m_{tot}, m_{Wb})$, where $m_{Wb} \equiv m(W_\nottop b_\nottop)$ and
$m_{tot} \equiv m(W_\nottop b_\nottop W_t b_t)$.
%%
%%%%%%%%%%%%%%%%%%%%%%%%%%%%%%%%%%%%%%%%%%%%%%%%%%%%%%%%%%%%%%%%%%%%%%%
\begin{figure}[tb!]
\begin{center}
\includegraphics[width=0.6\textwidth,clip,angle=0]{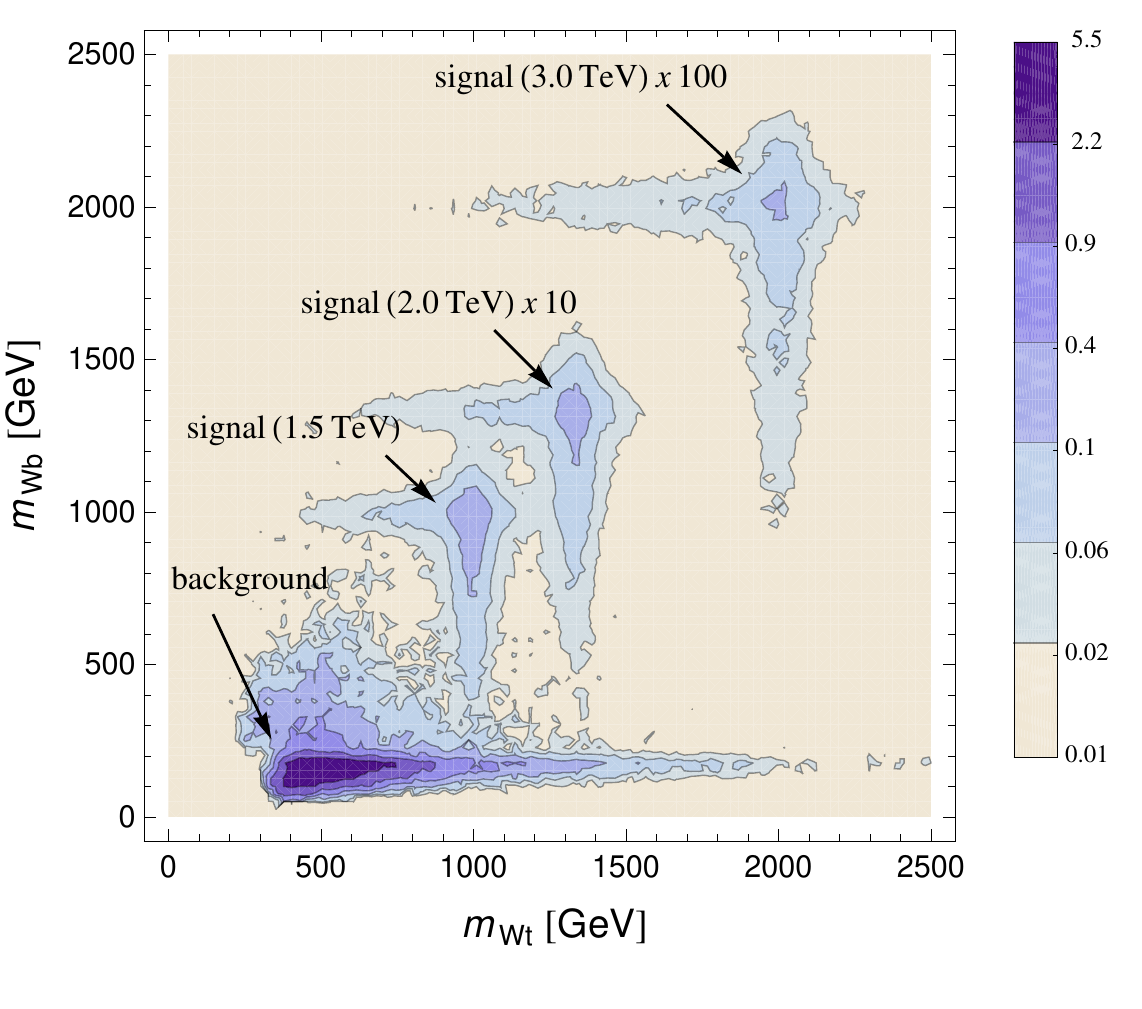}
\caption[]{
\label{fig:mWtmWb}
\small
Contour plot of the signal and total background cross sections after the cuts at `zero cost' of eq.(\ref{eq:zerocost7TeV}) in the plane 
$(m_{Wt}, m_{Wb})$, where $m_{Wb} \equiv m(W_\nottop b_\nottop)$ and $m_{Wt}\equiv m(W_\nottop W_t b_t)$. The center-of-mass energy 
is $\sqrt{s}=7\,$TeV. Different colors correspond to areas of different values of the doubly differential cross section 
$d^2\sigma/dm_{Wb}dm_{tot}$,  as reported in the vertical color key on the right in units $\text{ab}/\gev^2$.
The signals at $M_{G^*} =2.0\,$TeV and $3.0\,$TeV have been rescaled respectively by a factor 10 and 100. 
}
\end{center}
\end{figure}
%%%%%%%%%%%%%%%%%%%%%%%%%%%%%%%%%%%%%%%%%%%%%%%%%%%%%%%%%%%%%%%%%%%%%%%
%%
By simple inspection, one can see that the SM $t\bar t$ background can be strongly reduced by 
cutting at the same time on $m_{tot}$ and on $m_{Wb}$. We find that an additional cut on $m_{Wt}\equiv m(W_\nottop W_t b_t)$ 
is useful at large $M_{G^*}$ to further reduce the background and increase the signal significance, see Fig.~\ref{fig:mWtmWb}.
Although cutting on $m_{Wb}$  removes almost entirely the $t\bar t$ component of the signal,  this can  still be detected by
adopting the strategies proposed in previous analyses.  In the following, instead, we will focus on the  $G^* \to \tilde T t + Bb$ signal.

For each value of $M_{G^*}$ we find a set of optimized cuts that minimizes the integrated luminosity needed for a 
$5\sigma$ discovery.~\footnote{\label{ftn:5sigma}
We define the discovery luminosity to be the integrated luminosity for which a goodness-of-fit test of the SM-only hypothesis with Poisson 
distribution gives a  $\text{p-value}=2.85\times 10^{-7}$, which corresponds to a $5\sigma$ significance in the limit of a gaussian distribution.
If however this value is less than the luminosity for which
the total  (signal plus background) number of expected events is equal to 10,
this latter value is conservatively defined as the discovery luminosity.
}
These are:
\begin{equation} \label{eq:optimized}
\begin{aligned}
& \underline{M_{G^*} = 1.5\,\tev}: && \quad m_{tot} \geq 1300\,\gev && \quad m_{Wb}\geq 400\,\gev && \\[0.4cm]
& \underline{M_{G^*} = 2.0\,\tev}: && \quad m_{tot} \geq 1700\,\gev && \quad m_{Wb}\geq 600\,\gev && \\[0.4cm]
& \underline{M_{G^*} = 3.0\,\tev}: && \quad m_{tot} \geq 2500\,\gev && \quad m_{Wb}\geq 600\,\gev && \quad m_{Wt} \geq 700\,\gev \\[0.4cm]
& \underline{M_{G^*} = 4.0\,\tev}: && \quad m_{tot} \geq 3200\,\gev && \quad m_{Wb}\geq 700\,\gev && \quad m_{Wt} \geq 900\,\gev \, .
\end{aligned}
\end{equation}
In each case, the efficiency of the cut on $m_{Wb}$ \textit{after} imposing that on $m_{tot}$ is of the order of a few percent for the $WWbb$
background, and $\sim 90\%$ for the $G^* \to \tilde T t + Bb$ signal.
The values of the cross sections after these optimized cuts are reported in Tables~\ref{tab:cutflow7TeV2} and~\ref{tab:cutflow14TeV2} 
(in the columns labeled as `\textsc{opt}') respectively for $\sqrt{s}=7\,$TeV and $14\,$TeV.
%%
%%%%%%%%%%%%%%%%%%%%%%%%%%%%%%%%%%%%%%%%%%%%%%%%%
\begin{table}
\begin{center}
{\small
\begin{tabular}{|l|ll|ll|l|}
\hline 
 & & & & &  \\[-0.35cm]
\multirow{2}{*}{\textsf{LHC $\mathsf{7\,}$TeV}} & \multicolumn{2}{l|}{$M_{G^*} = 1.5\,$TeV} & \multicolumn{2}{l|}{$M_{G^*} = 2.0\,$TeV} 
                                                 &$M_{G^*} = 3.0\,$TeV \\[0.1cm]
& \textsc{opt} & \textsc{opt(ii)} & \textsc{opt} & \textsc{opt(ii)} & \textsc{opt}  \\[0.1cm]
\hline
& & & & &  \\[-0.3cm]
$G^*\to\tilde Tt + Bb$ 
                          & $20.0(1)$ & $14.9(1)$ & $2.28(1)$ & $1.85(1)$ & $0.0301(2)$ \\[0.25cm]
$WWbb$            & 0.17(6)   & 0.06(4)  & 0.06(4)   & 0.02(3)   & $<0.03$ \\[0.15cm]
$Wbbj$              & 0.32(2)   & 0.13(1)  & 0.044(7) & 0.022(5) & $0.003(2)$ \\[0.15cm]   
$Wbbjj$             & 0.11(2)   & 0.06(1)  & 0.023(8) & 0.008(5) & $0.003(4)$  \\[0.15cm] 
$W3j$                & 0.082(3) & 0.036(2)& 0.018(2) & 0.008(1) & $0.0009(3)$ \\[0.15cm]
$W4j$                & 0.039(3) & 0.015(2)& 0.011(1) & 0.005(1) & $0.0004(4)$ \\[0.15cm]
Total & & & & & \\
background      & $0.72(7)$ & $0.29(4)$ & $0.15(4)$ & $0.06(3)$ & $0.007^{+0.03}_{-0.004}$\\[0.1cm]
\hline
\end{tabular}
\caption{
\label{tab:cutflow7TeV2}
\small 
Cross sections, in fb, at $\sqrt{s}=7\,$TeV for the signal and the main backgrounds after imposing the 
cuts of eqs.(\ref{eq:acceptance}),(\ref{eq:zerocost7TeV}), and the optimized cuts of  eq.(\ref{eq:optimized}) (columns labeled as `\textsc{opt}').
In the case $M_{G^*} = 1.5\,$TeV and $2\,$TeV the columns labeled as `\textsc{opt(ii)}' report the value of the cross section
after the alternative set of optimized cuts of eq.(\ref{eq:optimizedB}), in addition to those of eqs.(\ref{eq:acceptance}),(\ref{eq:zerocost7TeV}).
For each channel, the proper branching fraction to a one-lepton final state has been included.
Details on how the statistical errors and upper limits on the cross sections have been computed and combined are given in Appendix~\ref{app:errors}.
}}
\end{center}
\end{table}
%%%%%%%%%%%%%%%%%%%%%%%%%%%%%%%%%%%%%%%%%%%%%%%%%
%%%%%%%%%%%%%%%%%%%%%%%%%%%%%%%%%%%%%%%%%%%%%%%%%
\begin{table}
\begin{center}
{\small
\begin{tabular}{|l|ll|ll|l|l|}
\hline 
 & & & & & & \\[-0.35cm]
\multirow{2}{*}{\textsf{LHC $\mathsf{14\,}$TeV}} & \multicolumn{2}{l|}{$M_{G^*} = 1.5\,$TeV} & \multicolumn{2}{l|}{$M_{G^*} = 2.0\,$TeV} 
                                                 &$M_{G^*} = 3.0\,$TeV & $M_{G^*} = 4.0\,$TeV \\[0.1cm]
& \textsc{opt} & \textsc{opt(ii)} & \textsc{opt} & \textsc{opt(ii)} & \textsc{opt}  & \textsc{opt} \\[0.1cm]
\hline
& & & & & & \\[-0.3cm]
$G^*\to\tilde Tt + Bb$ 
                          &  175.7(8) & 133.0(7) & 39.5(2) & 32.3(2)   & 2.76(1)         & 0.231(1) \\[0.25cm]
$WWbb$            &  3.9(3)     &  0.9(2)    & 1.0(2)    & 0.13(7)   & $0.02(5)$& $<0.04$ \\[0.15cm]
$Wbbj$              &  2.8(1)     &  1.34(8)  & 0.76(6)  & 0.37(4)   & 0.06(2)        & 0.005(7)  \\[0.15cm]   
$Wbbjj$             &  1.20(4)   &  0.59(3)  & 0.32(2)  & 0.16(1)   & 0.028(6)      & 0.002(2)  \\[0.15cm] 
$Wbbjjj$            &  0.10(1)   &  0.037(8)& 0.022(6)& 0.012(5) & $<0.002$   & $<0.002$\\[0.15cm] 
$W3j$                &  0.60(2)   &  0.26(1)  & 0.21(1)  & 0.095(8) & 0.043(5)      & 0.009(2)  \\[0.15cm]
$W4j$                &  0.28(2)   &  0.16(1)  & 0.10(1)  & 0.055(7) & 0.025(5)      & 0.003(1)   \\[0.15cm]
Total & & & & & & \\
background       & 8.9(3) & 3.3(2) & 2.4(2) & 0.82(8) & $0.18(5)$ & $(0.019^{+0.04}_{-0.007})$\\[0.1cm]
\hline
\end{tabular}
\caption{
\label{tab:cutflow14TeV2}
\small 
Cross sections, in fb, at $\sqrt{s}=14\,$TeV for the signal and the main backgrounds after imposing the 
cuts of eqs.(\ref{eq:acceptance}),(\ref{eq:zerocost14TeV}), and the optimized cuts of  eq.(\ref{eq:optimized}) (columns labeled as `\textsc{opt}').
In the case $M_{G^*} = 1.5\,$TeV and $2\,$TeV the columns labeled as `\textsc{opt(ii)}' report the value of the cross section
after the alternative set of optimized cuts of eq.(\ref{eq:optimizedB}), in addition to those of eqs.(\ref{eq:acceptance}),(\ref{eq:zerocost14TeV}).
For each channel, the proper branching fraction to a one-lepton final state has been included.
Details on how the statistical errors and upper limits on the cross sections have been computed and combined are given in Appendix~\ref{app:errors}. 
}}
\end{center}
\end{table}
%%%%%%%%%%%%%%%%%%%%%%%%%%%%%%%%%%%%%%%%%%%%%%%%%
%
The values of the corresponding discovery luminosity are reported in Table~\ref{tab:Ldisc7TeV}.
%%%%%%%%%%%%%%%%%%%%%%%%%%%%%%%%%%%%%%%%%%%%%%%%%
\begin{table}
\begin{center}
{\small
\begin{tabular}{r|cccc}
\multirow{2}{*}{\textsf{LHC $\mathsf{\sqrt{s} = 7\,}$TeV}} & \multicolumn{3}{c}{$M_{G^*} [\text{TeV}]$} & \\[0.15cm]
  & 1.5 & 2.0 & 3.0 & \\[0.1cm]
\cline{1-4}
& & & & \\[-0.3cm]
$L_{disc} \,[\text{fb}^{-1}]$ & 0.48 & 4.1 & $1.3 \cdot 10^3$ &
\\
\multicolumn{4}{c}{}
\\[0.7cm]
\multirow{2}{*}{\textsf{LHC $\mathsf{\sqrt{s} = 14\,}$TeV}} & \multicolumn{4}{c}{$M_{G^*} [\text{TeV}]$} \\[0.15cm]
 & 1.5 & 2.0 & 3.0 & 4.0 \\[0.1cm]
\hline
& & & & \\[-0.3cm]
$L_{disc} \,[\text{fb}^{-1}]$ & 0.054 & 0.24 & 3.4 & 57
\end{tabular}
\vspace{0.5cm}
\caption{
\label{tab:Ldisc7TeV}
\small 
Value of the integrated luminosity required for a $5\sigma$ discovery after the optimized cuts of eq.(\ref{eq:optimized}), 
for $\sqrt{s}=7\,$TeV (upper panel) and $\sqrt{s}=14\,$TeV (lower panel).
The $5\sigma$ discovery luminosity has been computed as explained in footnote~\ref{ftn:5sigma}.
}}
\end{center}
\end{table}
%%%%%%%%%%%%%%%%%%%%%%%%%%%%%%%%%%%%%%%%%%%%%%%%%

In the case of $M_{G^*} =1.5, 2.0\,$TeV, we found that imposing  cuts slightly stronger than those of eq.(\ref{eq:optimized}) can lead to 
a higher signal over background ratio, $\Scal/\Bcal$, (although the corresponding  discovery luminosity is also higher).
Specifically, we find that the set of cuts that maximizes $\Scal/\Bcal$ is:
\begin{equation} \label{eq:optimizedB}
\begin{aligned}
& \underline{M_{G^*} = 1.5\,\tev}: && \quad m_{tot} \geq 1300\,\gev && \quad m_{Wb}\geq 500\,\gev && \quad m_{Wt} \geq 900\,\gev \\[0.4cm]
& \underline{M_{G^*} = 2.0\,\tev}: && \quad m_{tot} \geq 1700\,\gev && \quad m_{Wb}\geq 600\,\gev && \quad m_{Wt} \geq 1100\,\gev \, .
\end{aligned}
\end{equation}
The values of the cross sections after imposing this alternative set of optimized cuts are reported in the columns of 
Tables~\ref{tab:cutflow7TeV2} and~\ref{tab:cutflow14TeV2} labeled as `\textsc{opt(ii)}', 
respectively for $\sqrt{s}=7\,$TeV and $14\,$TeV.

\subsection{Discovery reach on the parameter space}

\begin{figure}[tbp]
\begin{center}
\includegraphics[width=0.45\textwidth,clip,angle=0]{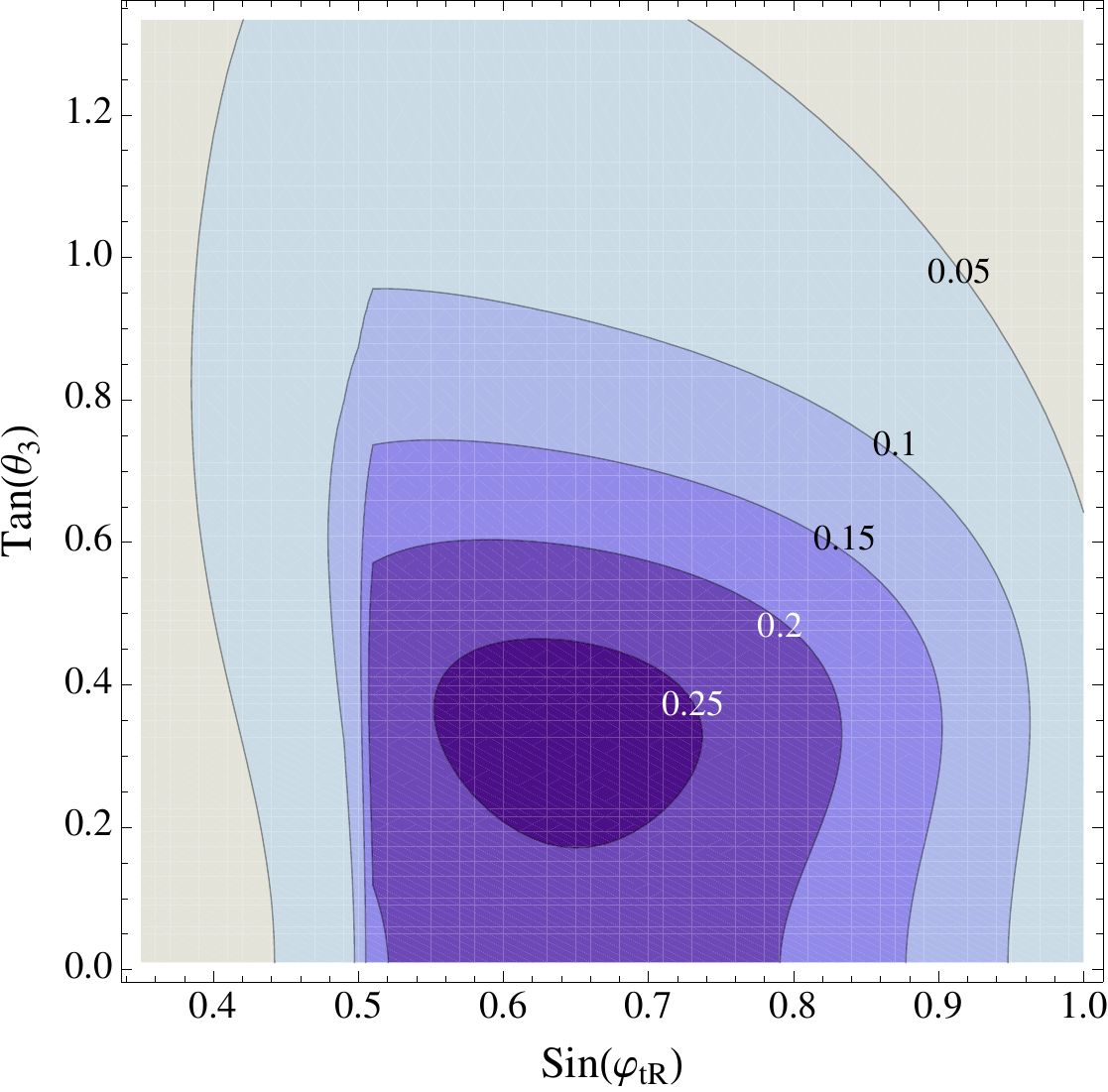}
\caption[]{
\label{fig:BRGstarWtb}
\small
Isocurves of constant branching ratio $BR(G^* \to \tilde T t+ Bb \to Wtb)$ in the plane $(\sin\varphi_{t_R}, \tan\theta_3)$ with
$M_{G^*}/m_{\tilde T}=1.5$ and $Y_* =3$.
}
\end{center}
\end{figure}

\begin{figure}[tbp]
\begin{center}
\includegraphics[width=0.70\textwidth,clip,angle=0]{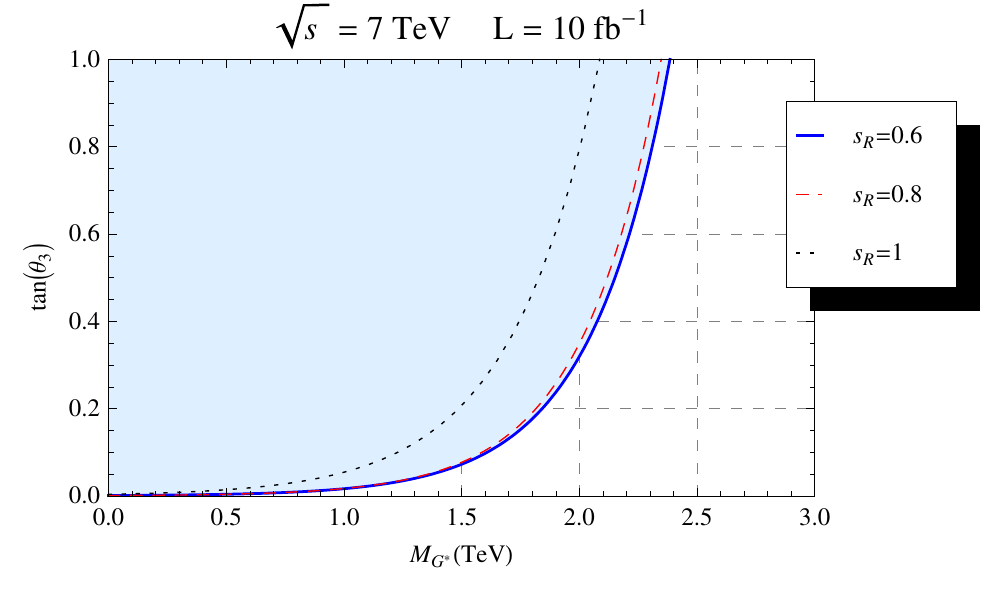}
\\[0.5cm]
\includegraphics[width=0.70\textwidth,clip,angle=0]{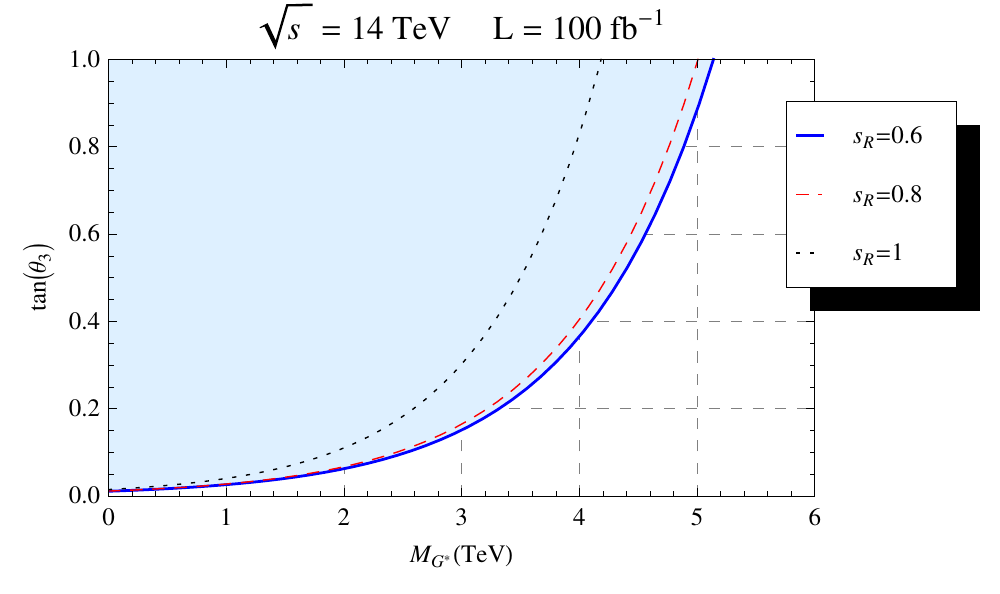}
\caption[]{
\label{fig:reach}
\small
LHC discovery reach in the plane $(M_{G^*},\tan\theta_3)$. The blue area shows the region where a  discovery of the 
signal $pp\to G^*\to \tilde Tt + Bb \to Wtb$ is possible at $5\sigma$ with $\sin\varphi_{t_R} =0.6$, $M_{G^*}/m_{\tilde T}=1.5$  and $Y_* =3$.
The reach at $\sin\varphi_{t_R} =0.8$ and  $\sin\varphi_{t_R} =1$ is shown respectively by the dashed red curve and the dotted black curve.
Upper plot: LHC at $\sqrt{s}=7\,$TeV with an integrated luminosity $L = 10\,\text{fb}^{-1}$; Lower plot:
 LHC at $\sqrt{s}=14\,$TeV with  $L = 100\,\text{fb}^{-1}$.
}
\end{center}
\end{figure}

All the numbers shown in Tables~\ref{tab:cutflow7TeV2}, \ref{tab:cutflow14TeV2} and~\ref{tab:Ldisc7TeV} hold for the TS5 model 
at the benchmark point of eq.(\ref{eq:benchmark}), where $BR(G^* \to \tilde T t+ Bb \to Wtb) = 0.25$ and the $G^*$ production cross section
is that of Fig.~\ref{fig:Gstarxsec}.
It is however legitimate to ask how these results change when varying the model's parameters.
The production cross section scales with $(\tan\theta_3)^2$, while
the branching ratio $BR(G^* \to \tilde T t+ Bb \to Wtb)$ is controlled by $\tan\theta_3$, $\sin\varphi_{tR}$, $Y_*$ and the ratio of heavy masses
$M_{G^*}/m_{\tilde T}$. 
To simplify the picture, we can fix the latter two parameters to the values adopted in the analysis, 
$M_{G^*}/m_{\tilde T}=1.5$  and $Y_* =3$, and  study the dependence on $\tan\theta_3$ and $\sin\varphi_{tR}$.
Figure~\ref{fig:BRGstarWtb} shows how the branching ratio varies  in the plane $(\sin\varphi_{tR},\tan\theta_3)$.
%%
%%%%%%%%%%%%%%%%%%%%%%%%%%%%%%%%%%%%%%%%%%%%%%%%%%%%%%%%%%%%%%%%%%%%%%%
%%%%%%%%%%%%%%%%%%%%%%%%%%%%%%%%%%%%%%%%%%%%%%%%%%%%%%%%%%%%%%%%%%%%%%%
%%
It is possible to estimate how the LHC discovery reach
varies with $\tan\theta_3$ and  $\sin\varphi_{tR}$ by simply rescaling the numbers in Tables~\ref{tab:cutflow7TeV2}, 
\ref{tab:cutflow14TeV2} to take into account the change in the production cross section and in the branching ratio to the final state $Wtb$.
The result is reported in Fig.~\ref{fig:reach}. The two plots show the 
region  in the plane $(M_{G^*},\tan\theta_3)$ where a $5\sigma$ discovery is possible for the LHC at $\sqrt{s}=7\,$TeV with $L = 10\,\text{fb}^{-1}$ 
(upper plot), and at $\sqrt{s}=14\,$TeV  with $L = 100\,\text{fb}^{-1}$  (lower plot).
Three different degrees of compositeness of the right-handed top quark are considered: $\sin\varphi_{tR} =0.6,0.8,1$.
%
%%
%%%%%%%%%%%%%%%%%%%%%%%%%%%%%%%%%%%%%%%%%%%%%%%%%%%%%%%%%%%%%%%%%%%%%%%
%%%%%%%%%%%%%%%%%%%%%%%%%%%%%%%%%%%%%%%%%%%%%%%%%%%%%%%%%%%%%%%%%%%%%%%
%%
We expect that this simple rescaling reproduces reasonably well
the actual reach one would obtain by optimizing the cuts at each different point of the parameter space.
A possible bias can arise in the limit in which
the decay width of the $G^*$ or those of the heavy fermions become
large, and tight cuts on the invariant masses have been applied. In this case, the efficiency of the kinematic
cuts will  in general depend non-trivially on the parameters via the particles' decay widths, and will not be
reproduced by a simple rescaling. 
In our analysis the cuts on $m_{Wb}$ and $m_{Wt}$ are always much below the heavy fermions' masses, and also 
the cut on the total invariant mass should be sufficiently below the $G^*$ mass to neglect, in first approximation,
any deviation from the case of a simple rescaling.

The final results of our analysis, summarized by Tables~\ref{tab:cutflow7TeV2}, \ref{tab:cutflow14TeV2} and~\ref{tab:Ldisc7TeV} and by 
the plots of Fig.~\ref{fig:reach}, are very encouraging. 
For example,  if $\sin\varphi_{tR} =0.6$ (a value which almost maximizes the branching ratio to $Wtb$ 
in our model, see Fig.~\ref{fig:BRGstarWtb}),  the LHC  at $\sqrt{s}=7\,$TeV and with $10\,$fb$^{-1}$ of integrated luminosity 
should be able to discover a $G^*$ with mass in the range $M_{G^*} = (1.8 - 2.2)\,$TeV  for $\tan\theta_3 = 0.2-0.5$. 
On the other hand, by running at the design c.o.m. energy $\sqrt{s}=14\,$TeV, the LHC discovery reach extends to the mass range
$M_{G^*} = (3.3 - 4.4)\,$TeV  for $\tan\theta_3 = 0.2-0.5$ with an integrated  luminosity $L =100\,$fb$^{-1}$.

%%%%%%%%%%%%%%%%%%%%%%%%%%%%%%%%%%%%%%
\section{Discussion}
\label{sec:conclusions}
%%%%%%%%%%%%%%%%%%%%%%%%%%%%%%%%%%%%%%

In order to better quantify the potentiality of our analysis,
it is useful to compare our results with those obtained in previous studies. 

Ref.~\cite{Agashe:2006hk}  searches for a KK-gluon in the channel $G^* \to t\bar t$ and makes use of a LR polarization asymmetry to 
enhance the signal significance after cutting on the total invariant mass.
The authors consider the case in which the right-handed top quark is fully composite, and 
adopt the following benchmark values of parameters: $\tan\theta_3 =0.2$, $\sin\varphi_{tR} =1$, $\sin\varphi_{L} =0.33$, $Y_* =3$.
The corresponding value of the couplings of $G^*$ to the light quarks, top quark and bottom quarks is:
$g_{G^*q\bar q} = g_{G^*b_R b_R} = -0.2 \, g_3$,  $g_{G^*t_L t_L} = g_{G^*b_L b_L} \simeq  g_3$,  $g_{G^*t_R t_R} = 5 \, g_3$.
The authors then assume that only decays of $G^*$ to pairs
of SM quarks are kinematically allowed, and obtain in this way a branching ratio $BR(G^* \to t\bar t)\simeq 0.95$.
At the end of their analysis, they find 
$\Scal/\Bcal =2 \, (1.6)$ and $\Scal/\sqrt{\Bcal} =11 \, (4.2)$ for $M_{G^*} = 3\, (4)\,$TeV, where $\Scal$ ($\Bcal$) 
is the number of signal (background)
events expected at $\sqrt{s}=14\,$TeV with an integrated luminosity $L =100\,\text{fb}^{-1}$.
These numbers are obtained without making use of the LR polarization asymmetry.
This observable was proposed in Ref.~\cite{Agashe:2006hk} (see also Ref.~\cite{Lillie:2007yh}) as a way to better distinguish the signal from 
the background,  but its quantitative impact  on the signal significance was not derived by the authors.
One way to compare with the above results is to consider the signal 
yield that we predict for $\tan\theta_3 = 0.2$ and rescale it
by a factor $0.95/BR(G^* \to \tilde T t + Bb \to Wtb)$, so that both results are normalized in the same way
(\textit{i.e.} to the same production cross section and decay branching ratio).
In this way we obtain $\Scal/\Bcal =9.3 \, (3.0)$,  $\Scal/\sqrt{\Bcal} =44 \, (7.3)$ for $M_{G^*} = 3\, (4)\,$TeV
with $L =100\,\text{fb}^{-1}$ at $\sqrt{s}=14\,$TeV.~\footnote{In order to be  conservative in our estimate,
we have included a $1\sigma$ upward fluctuation of the background from its central value quoted in Table~\ref{tab:cutflow14TeV2}.} 
This shows how powerful  a search for $G^* \to \tilde T t + Bb$ is, as a result of  the strong reduction of the SM background
obtained by cutting on the invariant mass of the heavy fermion,
compared to the `standard'  $t\bar t$ channel, although it must be noticed that the
sensitivity of the latter is expected to increase when the information on the polarization of the $t\bar t$ pair is included.
Perhaps, a  more sensible way to compare our results with those of Ref.~\cite{Agashe:2006hk} is by rescaling 
the $t\bar t$ signal yield obtained in the latter to the branching ratio $BR(G^* \to t\bar t)$ predicted in our model at $M_{G^*}/m_{\tilde T} =1.5$.
For example we consider the benchmark values of the right plot of Fig.~\ref{fig:hldecays2} 
($\tan\theta_3 =0.2$, $\sin\varphi_{tR} =0.8$, $\sin\varphi_{L} =0.41$, $Y_* =3$), for which the branching ratios to $Wtb$ and $t\bar t$
are comparable: $BR(G^* \to \tilde T t + Bb \to Wtb) = 0.21$, $BR(G^* \to t\bar t)\simeq 0.41$. After the appropriate rescaling,
we find that with $L =100\,\text{fb}^{-1}$  at $\sqrt{s}=14\,$TeV the analysis of Ref.~\cite{Agashe:2006hk} in this case predicts 
$\Scal/\Bcal =0.86 \, (0.69)$,  $\Scal/\sqrt{\Bcal} =4.7 \, (1.8)$ for $M_{G^*} = 3\, (4)\,$TeV, while our result reads $\Scal/\Bcal =2.0 \, (0.66)$,  
$\Scal/\sqrt{\Bcal} =9.7 \, (1.6)$.
On the other hand, in the extreme benchmark point of Ref.~\cite{Agashe:2006hk}, where  $t_R$ is fully composite, we have checked that the $t\bar t$
channel leads to a much better discovery reach than the one predicted by  our analysis based on the $Wtb$ channel.
An opposite result is instead obtained for the benchmark values of eq.(\ref{eq:benchmark}), for which the branching ratio to $t\bar t$ is very small.

To summarize, we find that as long as the right-handed top quark is not exactly fully composite and the heavy gluon can decay to one or two
heavy fermions, the branching ratio of the $t\bar t$ channel decreases and the analyses based on it become less powerful.
If decays to two heavy fermions are kinematically forbidden, on the other hand, a search based on the heavy-light decays to $Wtb$ is very much 
competitive for the discovery of  $G^*$ and  possibly stronger than the `standard' $t\bar t$  searches.
In addition, it gives the opportunity to discover both
the heavy gluon and the new fermions at the same time.
In fact, the process $G^* \to \tilde T t + Bb$ turns out to be  competitive for the discovery of the $\tilde T$ and $B$ as well.

It is useful, for example, to compare with the results obtained in Ref.~\cite{Mrazek:2009yu}, where the QCD pair production 
and the EW single production of the heavy bottom $B$ were studied. 
At the end of their analysis, the authors find $\Scal/\Bcal = 1.45$ and a discovery luminosity $L_{disc} = 4.3\,\text{fb}^{-1}$ for $M_B =1\,$TeV 
at $\sqrt{s}=14\,$TeV.  For the same value of the heavy bottom mass, $M_B =1\,$TeV, assuming $M_{G^*}/m_B = 1.5$, our analysis leads to a 
much smaller discovery luminosity, thanks to the larger rate of production of the signal.
Taking into account only the signal $G^* \to Bb \to Wtb$ (\textit{i.e.} rescaling the signal yields of 
Table~\ref{tab:cutflow14TeV2} by a factor $\sim 2/3$ to subtract  the contribution from $\tilde T$, see eq.(\ref{eq:benchmarkBR})),
we find  $\Scal/\Bcal = 5.1 - 30$ and a discovery luminosity $L_{disc} = (0.63 - 0.068)\,\text{fb}^{-1}$ if $\tan\theta_3$ varies in the range 
$0.2-0.5$.~\footnote{Notice that our estimate of the discovery luminosity is more conservative than the one performed in Ref.~\cite{Mrazek:2009yu}, 
which is based on $\Scal/\sqrt{\Bcal}$. For example, using the signal and background cross sections reported in Table~5 of
Ref.~\cite{Mrazek:2009yu} for  $M_B =1\,$TeV, we estimate $L_{disc} = 6.7\,\text{fb}^{-1}$, instead of $4.3\,\text{fb}^{-1}$.} 
This shows that the heavy-light decays of $G^*$ can be an important mechanism for the production of the top and bottom KKs
which should be taken into account, together with pair production and EW single production, to optimize the experimental searches.

Besides representing an additional and powerful way to discover the heavy fermions, the analysis of the heavy-light decays of $G^*$
also gives the opportunity to extract some important properties of the new fermionic sector. A thorough analysis of the various heavy-light
channels such as the one sketched in section~\ref{subsec:Gstarpheno} can indeed shed light on the quantum numbers of the heavy fermions and 
allow a determination of their couplings through the measurement of the branching ratios of  $G^*$. Finding a large coupling of the top partners 
to $G^*$, for example, may  give a first indication of their composite nature.

So far, the experimental searches for new heavy fermions have mostly concentrated on their QCD pair production. This is  expected to be the main 
production mechanism in the case of a fourth generation of chiral fermions, but it does not necessarily represent the best discovery process
in the case of new vector-like fermions.  At the same time, the experimental searches for a new heavy gluon have been so far performed
only in the dijet and $t\bar t$ channels. Our study shows that the presence of new heavy fermions has a strong impact on the phenomenology
of the heavy gluon, and opens up new possibilities of discovery via its decays to one SM and one heavy fermion.
The $t\bar t$ experimental searches that are based on a full reconstruction of a pair of top quarks are in fact blind to these new processes.
On the other hand, searches 
where the top quarks are not reconstructed, like for example those of Refs.~\cite{D05882,ATLAS087,ATLAS123}, 
are sensitive to our signal $G^* \to \tilde Tt + Bb \to Wtb$, but they are clearly less powerful than an optimized analysis, like the one
proposed in this work, which exploits the peculiar kinematics of the signal.
Optimizing the future experimental analyses to take into account the heavy-light decay topologies studied in this paper will
give the opportunity to better explore the parameter space of theories beyond the Standard Model and possibly lead to the discovery
of the sector responsible for the breaking of the electroweak symmetry.

%%%%%%%%%%%%%%%%%%%%%%%%%%%%%%%%%%%%%%
\section*{Note Added}
%%%%%%%%%%%%%%%%%%%%%%%%%%%%%%%%%%%%%%

While completing our work we became aware of the analysis of Ref.~\cite{Barcelo:2011wu}, where the
heavy-light decays of a KK gluon are also studied. The importance of these latter to explain the forward-backward 
top quark asymmetry at the Tevatron was originally discussed in Ref.~\cite{Barcelo:2011vk}, as it was pointed out to us by J.~Santiago.

%%%%%%%%%%%%%%%%%%%%%%%%%%%%%%%%%%%%%%
\section*{Acknowledgments}
%%%%%%%%%%%%%%%%%%%%%%%%%%%%%%%%%%%%%%

We thank B.~Mele for participation in the early stages of this work and 
 K.~Agashe, G.~Panico, A.~Wulzer for  discussions and comments.
The work of R.C. was partly supported by the ERC Advanced Grant No.~267985 
{\em Electroweak Symmetry Breaking, Flavour and  Dark Matter: One Solution for Three Mysteries (DaMeSyFla)}.
The work of N.V. was supported in part by DOE under contract number DE-FG02-01ER41155.

%%%%%%%%%%%%%%%%%%%%%%%%%%%%%%%%%%%%%%
\appendix
\section*{Appendix}
%%%%%%%%%%%%%%%%%%%%%%%%%%%%%%%%%%%%%%

\section{Lagrangian in the diagonal basis before EWSB}
\label{app:Lagrangian}

We collect here a few results on the TS5 model used  in section~\ref{sec:model}.
As explained in the text, the TS5 lagrangian (\ref{eq:Ltotal}) can be diagonalized, before EWSB, by performing a field rotation 
from the composite/elementary  to the mass-eigenstate basis. More details on such rotation will be given in a forthcoming
publication~\cite{wp}.  The diagonalized lagrangian reads
\begin{align}
\label{eq:Lrotated}
{\cal L} =& \, {\cal L}_{gauge}+ {\cal L}_{fermion}+ {\cal L}_{Higgs} \\[0.5cm]
\begin{split}   \label{eq:Lrotatedgauge}
{\cal L}_{gauge}= & -\frac{1}{4} G_{\mu\nu}G^{\mu\nu} \\[0.1cm]
 & +\frac{1}{2}\left(D_\mu G^*_\nu D_\nu G^*_\mu - D_\mu G^*_\nu D_\mu G^*_\nu \right)+\frac{1}{2}M^2_{G*}G^{* 2}_\mu 
     +\frac{ig_3}{2} \, G_{\mu\nu} \!\left[G^{*}_\mu,G^*_\nu \right] \\[0.1cm]
 & + 2i \, g_3 \cot 2\theta_3 \, D_\mu G^*_\nu \left[G^{*}_\mu,G^*_\nu \right]+\frac{g^2_3}{4}\left(\frac{\sin^4\!\theta_3}{\cos^2\!\theta_3} 
    + \frac{\cos^4\!\theta_3}{\sin^2\!\theta_3}\right) \left[G^{*}_\mu,G^*_\nu \right]^2 
\end{split}  \\[0.5cm]
\begin{split}  \label{eq:Lrotatedfermion}
{\cal L}_{fermion}= 
       & \, \bar{q} \, i\!\Dslash q+ \bar{\psi}\, i\!\Dslash\psi+\bar{\chi}\left( i\!\Dslash - m_\chi\right)\chi \\[0.1cm]
       & -g_3 \tan\theta_3 \, G^*_\mu \bar{q}\gamma^\mu q + g_3 \left( \sin^2\!\varphi_{\psi} \cot\theta_3 
           -  \cos^2\!\varphi_{\psi}\tan\theta_3\right) G^*_\mu \bar{\psi}  \gamma^\mu \psi \\[0.1cm]
       & + g_3 \, \frac{\sin\varphi_{\psi}\cos\varphi_{\psi}}{\sin\theta_3 \cos\theta_3} \, G^*_\mu \bar{\chi} \gamma^\mu \psi 
          +  g_3 \left( \cos^2\!\bar\varphi_{\chi} \cot\theta_3- \sin^2\!\bar\varphi_{\chi} \tan\theta_3 \right) G^*_\mu \bar{\chi} \gamma^\mu \chi \\[0.1cm]
       & + h.c. + O(s_2) 
\end{split} 
\\[0.5cm]
%\end{align}
%
%\begin{equation}
\begin{split}  \label{eq:LrotatedHiggs}
 {\cal L}_{Higgs}= 
  & \, |D_\mu H|^2 -V(H) \\[0.1cm]
  & + Y_*  \cos\varphi_L \cos\varphi_{tR} \, \bar{Q}_L\tilde{H} \tilde{T}_R + Y_* \cos\varphi_{tR} \, \bar{Q}_{uL}H\tilde{T}_R 
      -Y_*\sin\varphi_L\cos\varphi_{tR}\,\bar{q}_L\tilde{H}\tilde{T}_R \\[0.1cm]
  & - Y_*\sin\varphi_{tR} \,\bar{Q}_{uL} H t_R -Y_*\cos\varphi_L\sin\varphi_{tR}\,\bar{Q}_L\tilde{H}t_R 
     + Y_*\sin\varphi_L\sin\varphi_{tR}\, \bar{q}_L\tilde{H}t_R \\[0.1cm]
  & - Y_* \left( s_2 \sin\varphi_L  + s_3  \cos\varphi_L \right)\left(\cos\varphi_{tR}\, \bar{Q}^{'}_L\tilde{H}\tilde{T}_R 
     - \sin\varphi_{tR}\, \bar{Q}^{'}_L\tilde{H}t_R \right)\\[0.1cm]
  & + Y_*\, \bar{Q}_R \tilde{H} \tilde{T}_L +Y_*\, \bar{Q}_u H \tilde{T}_L - s_4 Y_*\, \bar{Q}^{'}_R\tilde{H} \tilde{T}_L \\[0.1cm]
  & + Y_*\cos\varphi_{bR} \, \bar{Q}_{dL}\tilde{H}\tilde{B}_R+ Y_*\cos\varphi_{bR} \, \bar{Q}^{'}_{L}H\tilde{B}_R 
      - Y_* \sin\varphi_{bR} \,\bar{Q}_{dL}\tilde{H}b_R\\[0.1cm]
  & - Y_*\sin\varphi_{bR} \,\bar{Q}^{'}_{L}H b_R - Y_* s_2\cos\varphi_{bR}\, \bar{q}_{L} H \tilde{B}_R + Y_* s_2 \sin\varphi_{bR}\,\bar{q}_{L} H b_R \\[0.1cm]
  & - Y_* s_3\sin\varphi_{bR} \, \bar{Q}_{L} H b_R + Y_* s_3\cos\varphi_{bR} \,\bar{Q}_{L} H \tilde{B}_R\\[0.1cm]
  & + Y_* \, Q^{'}_R H\tilde{B}_L+ Y_* \, \bar{Q}_{dR}\tilde{H}\tilde{B}_L + Y_* s_4 \, \bar{Q}_R H \tilde{B}_L + h.c.
\end{split}
%\end{equation}
\end{align}
%
%\\[0.2cm]
where $q = u,d,c,s$, $\psi = t_L, b_L , t_R, b_R$, and we have defined  $Q=(T,B)$, $Q'=(T',B')$, $Q_u=(T_{5/3}, T_{2/3})$, $Q_d=(B_{-1/3},B_{-4/3})$, 
$H=(\phi^{+}, \phi_0)$,  $\tilde{H}\equiv i\sigma^2 H^* = (\phi^\dag_0, -\phi^{-})$.  $\chi$ denotes any of the heavy fermions, except in the 
first term in the third line of eq.(\ref{eq:Lrotatedfermion}), where it denotes a top or bottom heavy partner, $T, B, \tilde T, \tilde B$.
Finally, $s_3$ and $s_4$ are defined as follows:
\begin{equation} \label{eq:s3s4}
s_3=\frac{\Delta_{L2}\bar{m}_{Q'}}{\Delta^2_{L1}+\bar{m}^2_Q-\bar{m}^2_{Q'}}\sin\varphi_{L}\, , \qquad
s_4=\frac{\Delta_{L2}\Delta_{L1}}{\Delta^2_{L1}+\bar{m}^2_Q-\bar{m}^2_{Q'}}\, .
\end{equation}
The coupling of $G^*$ to the fermions are thus the following (we neglect terms of $O(s_2)$):
\begin{align}
\label{eq:Gqq}
g_{G^* qq} =& -g_3 \tan\theta_3\, , && q = u,d,c,s \\[0.3cm]
\label{eq:Gpsipsi}
g_{G^* \psi \psi} =& 
  g_3 \left( \sin^2\!\varphi_{\psi} \cot\theta_3 -  \cos^2\!\varphi_{\psi}\tan\theta_3\right) \, , &&   \psi = t_L, b_L , t_R, b_R\\[0.3cm]
\label{eq:Ghl}
g_{G^* \chi\psi} =& 
  g_3 \, \frac{\sin\varphi_{\psi}\cos\varphi_{\psi}}{\sin\theta_3 \cos\theta_3}\, , && \chi\psi = Tt_L , Bb_L, \tilde T t_R, \tilde B b_R\\[0.3cm]
\label{eq:Ghh}
g_{G^* \chi\chi} =& 
  g_3 \, \left( \cos^2\!\bar\varphi_{\chi} \cot\theta_3- \sin^2\!\bar\varphi_{\chi} \tan\theta_3 \right)\, , && \chi = \text{any of the heavy fermions}\, ,
\end{align}
where $\sin\varphi_{tL} = \sin\varphi_{bL} \equiv \sin\varphi_{L}$, and we have defined $\sin\bar\varphi_{T_L} = \sin\bar\varphi_{B_L} \equiv \sin\varphi_{L}$,
$\sin\bar\varphi_{\tilde T_R} \equiv \sin\varphi_{tR}$, $\sin\bar\varphi_{\tilde B_R} \equiv \sin\varphi_{bR}$, and $\sin\bar\varphi_{\chi} = 0$ for
any other heavy fermion $\chi$. In particular, at the benchmark point of eq.(\ref{eq:benchmark}) we have: $g_{G^* qq} = -0.44\, g_3$, 
$g_{G^* t_R t_R} = 0.53\, g_3$,  $g_{G^* t_L t_L} = g_{G^* b_L b_L} = g_{G^* b_R b_R} = 0.40\, g_3$.
The decay rates of $G^*$ to two fermions are:
\begin{align}
\label{eq:GammaGqq}
\Gamma\left(G^{*}\rightarrow q\bar q \right) = &
  \,\frac{\alpha_{3}}{6}M_{G*} \tan^{2}\!\theta_{3}\, , \\[0.5cm]
\label{eq:GammaGpsipsi}
\Gamma\left(G^{*}\rightarrow\psi\bar{\psi} \right) = &
  \,\frac{\alpha_{3}}{12}M_{G*}\left(\sin^{2}\!\varphi_{\psi} \cot\theta_{3}-\cos^{2}\!\varphi_{\psi}\tan\theta_{3}\right)^{2}\, , \\[0.5cm]
\label{eq:GammaGpsichi}
 \Gamma\left(G^{*}\rightarrow\chi\bar{\psi}+\psi\bar{\chi}\right) =&
  \,\frac{\alpha_{3}}{6}M_{G*}\frac{\sin^{2}\!\varphi_{\psi} \cos^{2}\!\varphi_{\psi} }{\sin^{2}\theta_{3}\cos^{2}\theta_{3}}
  \left(1-\frac{m^{2}_{\chi}}{M^{2}_{G*}}\right)\left(1-\frac{1}{2}\frac{m^{2}_{\chi}}{M^{2}_{G*}}-\frac{1}{2}\frac{m^{4}_{\chi}}{M^{4}_{G*}}\right)\, , \\[0.5cm]
\label{eq:GammaGhh}
\begin{split}
\Gamma\left(G^{*}\rightarrow \chi\bar{\chi}\right) =& 
  \,\frac{\alpha_{3}}{12}M_{G*}\bigg\{ \left[\left(\cos^{2}\!\bar\varphi_{\chi} \cot\theta_{3}-\sin^{2}\!\bar\varphi_{\chi} \tan\theta_{3}\right)^{2}
  +\cot^{2}\!\theta_{3}\right] \left(1-\frac{m^{2}_{\chi}}{M^{2}_{G*}}\right)  \\
 & +6\left( \cos^{2}\!\bar\varphi_{\chi} \cot^2\!\theta_3 -\sin^{2}\!\bar\varphi_{\chi} \right)\frac{m^{2}_{\chi}}{M^{2}_{G*}}  \bigg\} 
 \sqrt{1-4\frac{m^{2}_{\chi}}{M^{2}_{G*}}}\,  .
\end{split}
\end{align}
Eq.(\ref{eq:GammaGqq}) reports the decay rate to a pair of each species of light fermions, $q = u,d,c,s$, and the contribution
of both chiralities has been included.
Similarly, eq.(\ref{eq:GammaGhh}) reports the decay rate to a pair of any of the heavy fermions, $\chi$, and the contribution of both chiralities 
has been included. As above, $\psi = t_L, b_L, t_R, b_R$ in eqs.(\ref{eq:GammaGpsipsi}),(\ref{eq:GammaGpsichi}).

The heavy fermions mostly decay to one longitudinally-polarized vector boson or Higgs boson plus one top or bottom quark, due to their
large Yukawa coupling. The decay rates into the three possible channels are:
\begin{align}
\label{eq:GHW}
\begin{split}
\Gamma\left(\chi\rightarrow W_L\psi\right) =& \,\frac{\lambda^2_{W\chi}}{32 \pi}M_{\chi}
\left[\left( 1+\frac{m^2_{\psi}-M^2_W}{M^2_{\chi}}\right)\left( 1+\frac{m^2_{\psi}+2M^2_W}{M^2_{\chi}}\right)-4\frac{m^2_{\psi}}{M^2_{\chi}}\right] \\
& \times\sqrt{1-2\frac{m^2_{\psi}+M^2_W}{M^2_{\chi}}+\frac{\left(m^2_{\psi}-M^2_W\right)^2 }{M^4_{\chi}}}
\end{split} \\[0.5cm]
\label{eq:GHZ}
\begin{split}
\Gamma\left(\chi\rightarrow Z_L\psi\right) =&\,\frac{\lambda^2_{Z\chi}}{64 \pi}M_{\chi}
\left[\left( 1+\frac{m^2_{\psi}-M^2_Z}{M^2_{\chi}}\right)\left( 1+\frac{m^2_{\psi}+2M^2_Z}{M^2_{\chi}}\right)-4\frac{m^2_{\psi}}{M^2_{\chi}}\right] \\
 &\times\sqrt{1-2\frac{m^2_{\psi}+M^2_Z}{M^2_{\chi}}+\frac{\left(m^2_{\psi}-M^2_Z\right)^2 }{M^4_{\chi}}} 
\end{split}\\[0.5cm]
\label{eq:GHh}
\Gamma\left(\chi\rightarrow h\psi\right)=& \,\frac{\lambda^2_{h\chi}}{64 \pi}M_{\chi}
\left( 1+\frac{m^2_{\psi}}{M^2_{\chi}}-\frac{M^2_{h}}{M^2_{\chi}}\right)
\sqrt{\left(1-\frac{m^2_{\psi}}{M^2_{\chi}}+\frac{M^2_{h}}{M^2_{\chi}}\right)^2-4\frac{M^4_{h}}{M^4_{\chi}}}\, .
\end{align}
At $O(r)$, by using the Equivalence Theorem~\cite{equivth},  the couplings $\lambda_{W\chi}$, $\lambda_{Z\chi}$, $\lambda_{h\chi}$ can be extracted 
from the coefficients of the Yukawa terms in the diagonalized lagrangian (\ref{eq:Lrotated}).~\footnote{See for example the discussion 
in footnote 2  of Ref.~\cite{Contino:2008hi}.} In the case of the top and bottom partners, one has:
\begin{gather}
 \lambda_{WT} = 0 \, , \qquad \lambda_{ZT}=\lambda_{hT}=Y_* \cos\varphi_L \sin\varphi_{tR} \\[0.4cm]
 \lambda_{WB}= Y_* \cos\varphi_L \sin\varphi_{tR}  \, ,\qquad  \lambda_{ZB} = \lambda_{hB} = 0 \\[0.4cm]
 \lambda_{W\tilde{T}}=\lambda_{Z\tilde{T}}=\lambda_{h\tilde{T}}= Y_* \sin\varphi_L \cos\varphi_{tR} \, , \\[0.4cm]
 \lambda_{W\tilde{B}} \simeq Y_* s_2 \cos\varphi_{bR} \, , \qquad \lambda_{Z\tilde{B}}=\lambda_{h\tilde{B}} \sim Y_* \sin\varphi_{bR} \cos\varphi_{bR} \times O(r)\, .
\end{gather}
Notice, in particular, that the decay of $\tilde B$  to $Wt$ is suppressed by a factor $s_2$, since it must proceed via the mixing of the elementary $b_L$
to the composite $B^\prime$. On the other hand, $\tilde B$ can decay to $Zb_R$, $hb_R$ at next-to-leading order in $r$.
Since under our assumptions $s_2/\sin\varphi_{bR} = s_2/\sin\varphi_{L} = m_b/m_t \ll r \lesssim 1$, it follows 
$BR(\tilde{B}\to Zb + hb )\gg BR(\tilde{B}\to Wt)$.  We thus neglect the decay $\tilde{B}\to Wt$ in our analysis.

\section{Statistical Errors}
\label{app:errors}

In this appendix we give some details on how the statistical errors that appear in
Tables~\ref{tab:cutflow7TeV2} and~\ref{tab:cutflow14TeV2}  are computed and combined.

Let $\sigma = \lambda \, a$ be the cross section for a given process, where $\lambda$ is the number of events and $a$
is a proportionality factor with no uncertainty. Then, if a Montecarlo simulation of the process returns $n$ events, the true value $\lambda$ is
estimated following a  Bayesian approach with flat prior and posterior Poisson probability $p(\lambda | n) = \lambda^n \exp(-\lambda)/n!$.
The latter has mean $E[\lambda] = n+1$ and variance $V[\lambda] = n+1$, which implies a standard deviation on the
cross section $\delta \sigma = \sigma/\sqrt{n+1}$. All the statistical errors on individual cross sections appearing in 
Tables~\ref{tab:cutflow7TeV2} and~\ref{tab:cutflow14TeV2} have been computed according to this formula.

In the case in which a set of kinematic cuts is applied to a sample of $n_0$ simulated events with  initial cross section $\sigma_0$, 
and no event passes the cuts, an upper bound on the final cross section can  be derived at a given confidence level $\alpha$ as the value $\lambda_*$
for which 
\begin{equation*}
\int_0^{\lambda_*} \!\! d\lambda \; p(\lambda | 0) =\alpha \, . 
\end{equation*}
It thus follows $\lambda_* = -\log (1-\alpha)$, which implies an upper limit
on the final cross section $\sigma < 1.1/(n_0+1)\, \sigma_0$ at $68\%$ CL.
All the upper limits that appear in Tables~\ref{tab:cutflow7TeV2} and~\ref{tab:cutflow14TeV2} are computed according to this
formula and are at $68\%$ CL.

When summing over the cross section of several processes, the statistical errors are combined in quadrature:
\begin{equation*}
\delta \sigma_{tot} = \sqrt{\sum_i (\delta\sigma_i)^2}.
\end{equation*}
In the case in which on one of the initial cross sections there is an upper bound, $\sigma_j < \bar\sigma_j$ at $68\%$ CL, 
an asymmetric error on the total cross section is derived 
%by treating the upper bound as an upper ward 
as follows:
\begin{equation*}
\delta \sigma_{tot}^+ = \sqrt{\sum_{i\not =j} (\delta\sigma_i)^2 + \bar \sigma_j^2} \, ,
\qquad \quad
\delta \sigma_{tot}^- = \sqrt{\sum_{i\not =j} (\delta\sigma_i)^2} \, .
\end{equation*}
The above definition is straightforwardly extended to the case in which more than one of the initial cross sections have an upper bound.

%%%%%%%%%%%%%%%%%%%%
%% References
%%%%%%%%%%%%%%%%%%%%

\end{document}